\documentclass [12pt]{article}
\usepackage{lscape}                                           %
\usepackage{amsmath}
\usepackage{amsfonts} 
\usepackage{color}
\usepackage{graphicx}
\usepackage{lscape}
\newcommand{\PaMoSpFull}{\{x_t,\alpha_t,f_t\}}
\newcommand{\PaMoSp}{\{x_t,\alpha_t\}}
\newcommand{\PaMoSpht}{\{x_{\hat t}, \alpha_{\hat t} \}}
\newcommand{\PaMoSpNDue}{\{0,\alpha_t; \infty, \alpha_{t+1}\}}
\newcommand{\PaMoSpNDuen}[1]{\{0,\alpha_{#1}; \infty, \alpha_{N}\}}
\newcommand{\PaMoSpNDuenn}[2]{\{0,\alpha_{#1}; \infty, \alpha_{#2}\}}

\newcommand{\xl}{{x_{loc}}}

\newcommand{\xii}[1]{{x_{#1}}}

\newcommand{\bou}{_{bou}}

\newcommand{\Xz}{X^z}
\newcommand{\Xbz}{X^{\bar z}}

\newcommand{\bep}{{\bar \epsilon}}

\newcommand{\bc}{{\bar c}}
\newcommand{\bk}{{\bar k}}
\newcommand{\bt}{{\bar t}}
\newcommand{\bu}{{\bar u}}
\newcommand{\bv}{{\bar v}}
\newcommand{\bx}{{\bar x}}
\newcommand{\bz}{{\bar z}}

\newcommand{\bX}{{\bar X}}
\newcommand{\bY}{{\bar Y}}

\newcommand{\cF}{{\cal F}}
\newcommand{\cG}{{\cal G}}
\newcommand{\cH}{{\cal H}}

\newcommand{\cN}{{\cal N}}

\newcommand{\cS}{{\cal S}}
\newcommand{\cT}{{\cal T}}

\newcommand{\cX}{{\cal X}}
\newcommand{\cbX}{{\bar{ \cal X}}}

\newcommand{\htt}{{\hat t}}
\newcommand{\hx}{{\hat x}}

\newcommand{\uno}{\mathbb{I}}

\newcommand{\Z}{\mathbb{Z}}

\newcommand{\R}{\mathbb{R}}

\newcommand{\C}{\mathbb{C}}

\newcommand{\du}{\partial_u}
\newcommand{\dbu}{{\bar\partial_{\bar u}}}

\newcommand{\ke}{{k_\epsilon}}
\newcommand{\kbe}{k_{\bar\epsilon}}
\newcommand{\kei}[1]{{k_{\epsilon_{#1}}}}
\newcommand{\kbei}[1]{{k_{\bar\epsilon_{#1}}}}

\newcommand{\vect}[2]{\left( \begin{array}{c} #1 \\ #2 \end{array} \right) }

\newcommand{\mycases}[2]{\left\{ \begin{array}{c} #1 \\ #2 \end{array} \right. }

\newcommand{\oh}{\frac{1}{2}}

\newcommand{\COMMENTO}[1]{}

\newcommand{\COMMENTOO}[1]{}

\begin{document}
\title{
Canonical quantization of a string describing $N$ branes at angles 
}

\author{
{Igor Pesando$^1$}
\\
~\\
~\\
$^1$Dipartimento di Fisica, Universit\`a di Torino\\
and I.N.F.N. - sezione di Torino \\
Via P. Giuria 1, I-10125 Torino, Italy\\
\vspace{0.3cm}
\\{ipesando@to.infn.it}
}

\maketitle
\thispagestyle{empty}

\abstract{
We study the canonical quantization of a bosonic string in presence of
$N$ twist fields.
This generalizes the quantization of the twisted string in two ways:
the in and out states are not necessarily twisted and the number of
twist fields $N$ can be bigger than $2$.

In order to quantize the theory we need to find the normal modes.
Then we need to define a product between two modes which is
conserved.
Because of this we need to use the Klein-Gordon product and
to separate the string coordinate into the classical and the
quantum part. The quantum part has different boundary conditions than
the original string coordinates 
but these boundary conditions are precisely those
which make the operator describing the equation of motion 
self adjoint. 

The splitting of  the string coordinates into a classical and quantum
part allows the formulation of an improved overlap principle.
Using this approach we then proceed in computing the generating
function for the generic  correlator with $L$ untwisted 
operators and $N$ (excited) twist fields for branes at angles.
We recover as expected the results previously obtained using the path integral.
This construction explains why these correlators 
are given by a generalization of the Wick theorem.
%
}
\\
\\
keywords: {D-branes, Conformal Field Theory}
\\
\\

\newpage


\section{Introduction and conclusions}
\COMMENTO{
Comment on Bogoliubov transformation, different vacua
}
Since their introduction, D-branes have been very important in the formal
development of string theory as well as in attempts to apply string
theory to particle phenomenology and cosmology. 
However, the requirement of chirality in any physically realistic
model  leads to a somewhat restricted number of possible D-brane
set-ups.  An important class is intersecting brane models where
chiral fermions can arise at the intersection of two branes at angles.
An important issue for these models is the computation of Yukawa
couplings and flavour changing neutral currents.

Besides the previous computations many  other computations 
often involve correlators of twist fields and excited
twist fields. 
It is therefore important and interesting in its own to be able to
compute these correlators.
As known in the literature \cite{Dixon:1986qv} 
and explicitly shown in \cite{Pesando:2014owa} for the branes at
angles case and in less precise way in \cite{Pesando:2011ce}
for the case of magnetized branes these computations boil down to the
knowledge of the Green function in presence of twist fields and of the
correlators of the plain twist fields.
In many previous papers correlators with excited twisted fields have been 
computed on a case by case basis without a clear global picture, see
for example
(\cite{Burwick:1990tu}, \cite{Erler:1992gt}, \cite{Anastasopoulos:2013sta}).

In this technical paper we have analyzed the $N$ excited twist fields
amplitudes with $L$ boundary vertices at tree
level for open strings localized at $D$-branes intersections on $R^2$
(or $T^2$) and we rederive the results of \cite{Pesando:2014owa}.

We will nevertheless follow a different approach from most of the
literature in a twofold way.
Firstly, we use the so called Reggeon vertex \cite{SDS}, 
which allows to compute the generating function
of all correlators, in particular we will use the formulation put
forward in \cite{Petersen:1988cf}.
Secondly we use the canonical quantization approach while all the
previous literature has used 
the classical path integral approach
(\cite{Dixon:1986qv},\cite{Atick:1987kd}).
In the case at hand the path integral approach is more
efficient than the also classical sewing approach
(\cite{Corrigan:1975sn}, \cite{DiBartolomeo:1990fw}).
This approach has been explored in many papers in the branes at
angles setup as well as the T dual magnetic branes setup see for example
(\cite{Bianchi:1991rd}, \cite{Antoniadis:1993jp},
\cite{Gava:1997jt}, 
\cite{David:2000um},
\cite{Abel:2003vv}, \cite{Cvetic:2003ch}, 
\cite{Lust:2004cx},
\cite{Bertolini:2005qh}, \cite{Lawrence:2007bk},
\cite{Duo:2007he},\cite{Choi:2007nb},
\cite{Conlon:2011jq},
\cite{Pesando:2012cx}).

\begin{figure}[hbt]
\begin{center}
\def\svgwidth{350px}
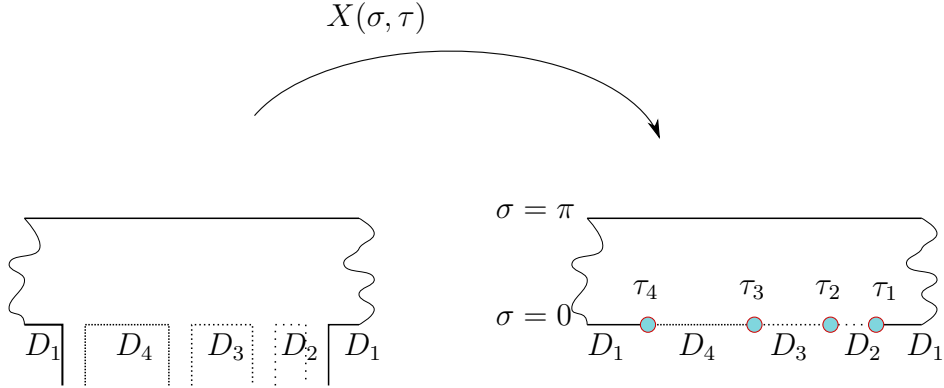
\end{center}
\vskip -0.5cm
\caption{The interacting strings are mapped into boundary condition discontinuities.
}
\label{fig:strings2points}
\end{figure}

At the heart of the path integral approach is the idea that the
interaction of a string with twisted strings in the fundamental state
can be replaced by a discontinuity on the string boundary conditions.
This is depicted in  figure \ref{fig:strings2points}.
We use this idea as the starting point of our computation based on
canonical formalism.
The fact that we have boundary conditions which change with the worldsheet time
implies that we have a very mild worldsheet time dependent worldsheet
metric. Hence the usual quantization cannot be applied in a
straightforward way but we have to find a proper way to defined the
product between modes. 
This is done using the Klein-Gordon product used in General Relativity.
To have a well defined, worldsheet time independent product between
two modes implies that we have to split the string into a classical and
quantum part and quantize the quantum part only, exactly as in the path
integral approach.
Since this procedure is at variance with the usual one we
check that we recover the standard results in the cases of the untwisted
string and of the usualm twisted ones.
All this is done in section \ref{setup_metric_consistency} using the
Klein-Gordon product used in General Relativity.
In this section we discuss also the expression of the Hamiltonian in
term of oscillators. We find that it is quadratic in oscillator 
but not diagonal since we have an almost free theory with worldsheet
time dependent background.
We derive also the modes for the three twists case. We are able to
find many orthogonal basis but all of them are missing of one mode with
respect to the basis of the in string.
This fact can be partially understood as the consequence of the
boundary ``interactions'' which break some symmetries of the original
in string. 
Since we have not understood this issue completely, we resort to using
an improved version of the standard overlap approach 
which is also used in quantum mechanics
in presence of discontinuities of the Hamiltonian.

In section \ref{Green_function_in_out_vacua} we tackle the problem of
computing the in and out vacua which necessarily differ since we have
a worldsheet time dependence in the boundary conditions. 
In principle it should be possible to compute them from the basic
principles. Because of the not complete understanding of modes we
compute the out vacuum as a kind of surface state (an exponential of
an expression  quadratic in the operators) of the in vacuum assuming
the knowledge of the Green function. This is however not a big issue since it
can be derived using the analytic properties and boundary conditions.

In section \ref{reggeons} we perform the actual computations of the
generating functions for amplitudes involving  plain and excited
twisted states. This is done in steps. First considering the
amplitudes with plain unexcited twisted fields and arbitrary untwisted
states. Then considering amplitudes with excited twisted states
without untwisted ones and finally, assembling all.

Our main result is to be able to rederive in a different way
the generating function of correlators with $N$ excited twists and 
$L$ untwisted states found in \cite{Pesando:2014owa}. 
It is given in
eq. (\ref{reggeon-excited-twists+bou}) which shows that all
correlators can be computed once the $N$ plain twist  operators
correlator  together with the Green function in presence of
these $N$ twists are known.
This expression  requires the precise knowledge of
the Green function \footnote{
Note that the Green functions used in this paper are dimensionful and
normalized as $\partial_u \bar\partial_\bu G^{I J}(u,\bu; v, \bv;\{
\epsilon_t\}) = -\frac{\alpha'}{2} \delta^{I J} \delta^2(u-v)$.}
 and its regularized versions. Luckily these are well known.
From these expressions it is clear that the computation of amplitudes,
i.e. moduli integrated correlators, with
(untwisted) states carrying momenta are very unwieldy because Green
functions can at best be expressed as sum of product of type D
Lauricella functions. This should however not be a complete surprise
since in \cite{Hamidi:1986vh} it was shown that twist fields
correlators in orbifold setup are connected to loop amplitudes which,
up to now, have not been expressed in term of simpler functions.
%

\section{The setup of branes at angles}
\label{setup_metric_consistency}
\begin{figure}[hbt]
\begin{center}
\def\svgwidth{350px}
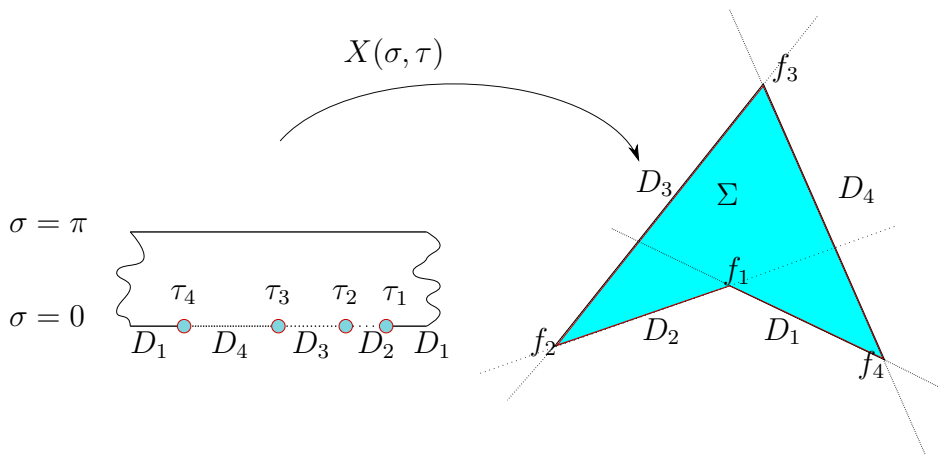
\end{center}
\vskip -0.5cm
\caption{Map from the Minkowskian worldsheet to the target polygon
  $\Sigma$ with a plain in and out string.
The map $X(\sigma,\tau)$ folds the $\sigma=0$ starting from $\tau=-\infty$ in a
counterclockwise direction.
}
\label{fig:stripe2polygon}
\end{figure}

\begin{figure}[hbt]
\begin{center}
\def\svgwidth{250px}
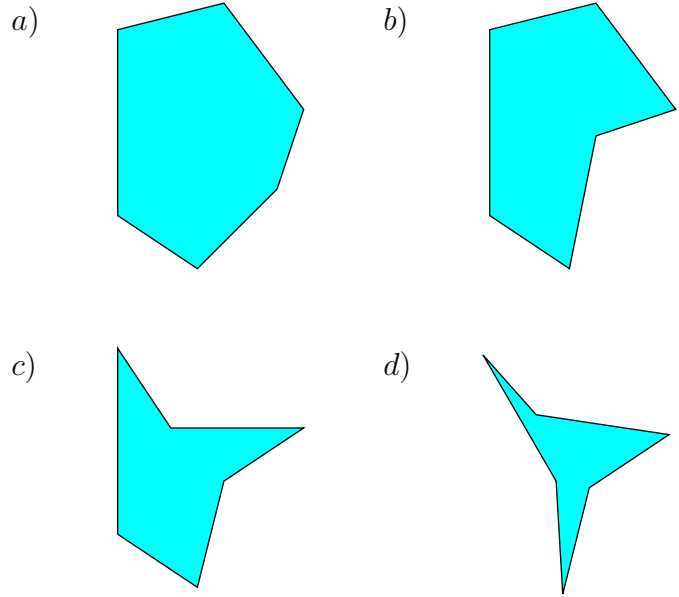
\end{center}
\vskip -0.5cm
\caption{The four different cases with $N=6$. 
$a)$ 
$M=4$ 
$b)$ 
$M=3$.
$c)$ 
$M=2$.
$d)$ 
$M=1$.
}
\label{fig:6gons}
\end{figure}

The Euclidean action for a string configuration is given by
\begin{equation}
S_E
=
\frac{1}{4\pi \alpha'} \int d\tau_E \int_0^\pi d\sigma~ (\partial_\alpha X^I)^2
=
\frac{1}{4\pi \alpha'} \int_H d^2 u~ 
(\du \Xz \dbu \Xbz + \dbu \Xz \du \Xbz )
\label{S_E}
\end{equation}
where $u=e^{\tau_E + i \sigma}\in H$, the upper half plane, 
$d^2 u= e^{2 \tau_E} d\tau_E d\sigma= \frac{d u ~d\bar  u}{2 i}$
and $I=1,2$ or $z, \bz$ so that $ \Xz= \frac{1}{\sqrt{2}}(X^1+i X^2)$, 
$ \Xbz= X^{z *}$.
The complex string coordinate is a map from the upper half plane to a
closed polygon $\Sigma$ in $\C$, i.e. $X:H \rightarrow \Sigma\subset \C$.
For example in fig. \ref{fig:stripe2polygon} we have pictured the interaction
of $N=4$ branes at angles $D_t$ with $t=1,\dots N$. The interaction
between brane $D_t$ and $D_{t+1}$ is in $f_t\in\C$. We use the rule
that index $t$ is defined modulo $N$.
As shown in \cite{Pesando:2011ce} given the number of twist fields
$N$ there are $N-2$ different sectors.
They are labeled by an integer $M$, $1\le M\le N-2$ 
which is in correspondence with the number of reflex angles (the
interior angles bigger than $\pi$), more precisely $M$ is $N-2$ minus the
number of reflex angles  as shown in
figure \ref{fig:6gons} in the case $N=6$.
The intuitive reason why they are different is that we need go through
the straight line, i.e. no twist,  if we want to go
from a reflex angles to a convex one.

\subsection{Splitting into classical and quantum part and Klein-Gordon metric}
\begin{figure}[hbt]
\begin{center}
\def\svgwidth{300px}
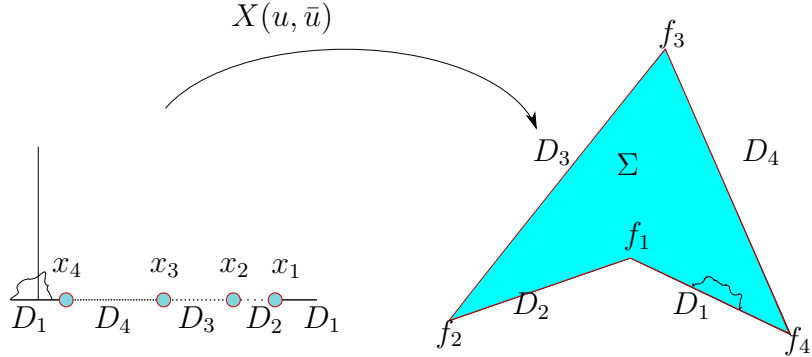
\end{center}
\vskip -0.5cm
\caption{Map from the upper half plane to the target polygon
  $\Sigma$ with untwisted in and out strings.
The map $X(u,\bu)$ folds the boundary of the upper half plane 
starting from $x=-\infty$ in a counterclockwise direction and
preserves the orientation.
}
\label{fig:H2polygon}
\end{figure}

\begin{figure}[hbt]
\begin{center}
\def\svgwidth{300px}
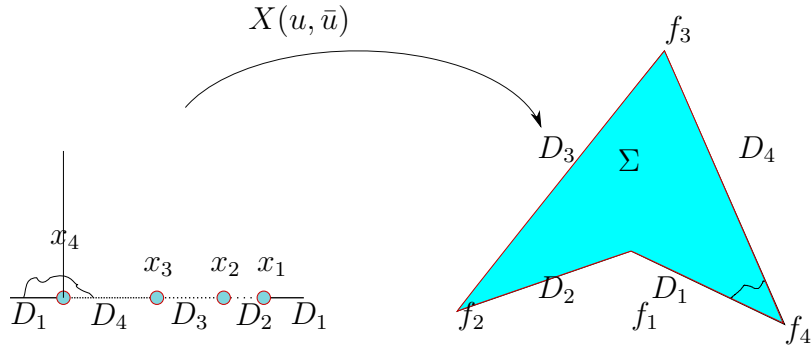
\end{center}
\vskip -0.5cm
\caption{Map from the upper half plane to the target polygon
  $\Sigma$ with a twisted in and an untwisted out string.}
\label{fig:H2polygon_twisted_in}
\end{figure}

In order to proceed with the canonical quantization
we want finding the normal modes associated with the equations of motion
\begin{equation}
\du\dbu \Xz = \du \dbu \Xbz =0
~~~~
u\in H
\label{eom}
\end{equation}
and the boundary conditions
\begin{align}
e^{- i \pi \alpha_t}
 \partial_y \Xz(u,\bu)|_{u=x+ i 0^+ } 
+
e^{ i \pi \alpha_t}
\partial_y \Xbz(u,\bu)|_{u=x+ i 0^+ } 
&= 0
~~~~
x_{t}<x<x_{t-1}
\nonumber\\
e^{-i \pi \alpha_t} \Xz(u,\bu)|_{u=x+ i 0^+ } 
-
e^{ i \pi \alpha_t} \Xbz(u,\bu)|_{u=x+ i 0^+ } 
&= 
2 i g_t
~~~~
x_{t}<x<x_{t-1}
.
\label{global-boundary-upper}
\end{align}
The previous constraints are simply stating that when
$x_{t}<x<x_{t-1}$ a boundary of the string is on the brane $D_t$.
The brane $D_t$ is described in a well adapted coordinate system as 
$\sqrt{2} i X^{2_t}=e^{-i \pi \alpha_t} \Xz - e^{ i \pi \alpha_t} \Xbz
= 2 i g_t\in i \R $,  i.e. it extends along $X^{1_t}$ with 
$\sqrt{2} X^{1_t}=e^{-i \pi \alpha_t} \Xz + e^{ i \pi \alpha_t} \Xbz$.
Therefore  the string has Dirichlet boundary condition in the
$X^{2_t}$ direction and  has Neumann boundary condition in the
perpendicular direction $X^{1_t}$.
In particular $\sqrt{2} |g_t|$ is the distance of the brane from the origin
and
\begin{equation}
f_t= \frac{e^{i \pi \alpha_{t+1}} g_t - e^{i \pi \alpha_{t}} g_{t+1}
}{ \sin \pi ( \alpha_{t+1} - \alpha_{t} )}
\label{f-intersection}
\end{equation}
is the intersection point between $D_t$ and $D_{t+1}$.
The configuration can be pictured as in figure \ref{fig:H2polygon}
in the Euclidean case. 
This configuration corresponds to the Minkowskian configuration of 
figure \ref{fig:stripe2polygon}
where both the incoming and the outgoing strings are untwisted. 
A similar configuration is drawn in figure
\ref{fig:H2polygon_twisted_in}. 
In this case the incoming string is twisted because one twist is
sitting in the origin and the outgoing is untwisted.
Obviously there is also a third possibility pictured in figure \ref{fig:H2polygon_twisted_in_out} where both the incoming
and outgoing strings are twisted, this corresponds to the case where
there is a twist at $x=0$ and one at $x=\infty$. 
When there are $N=2$ twists this is the usual twisted string.

The issue is now to find a (non positive definite) product for the
modes which is conserved in 
(Euclidean) time. This issue is less trivial than usual because our spacetime,
i.e. the worldsheet, is changing with (Euclidean) time even if in
a mild way through the change of the boundary conditions.
The solution to this problem is well known in General Relativity.
Since we are dealing with free fields satisfying the
Klein-Gordon equation we know that there is a conserved current.
Explicitly given any two solutions $F_i=(f_i^z, f_i^\bz)^T$ with $i=1,2$ the
current
\begin{equation}
j_\alpha= i~F_1^\dagger \stackrel{\leftrightarrow}{\partial_\alpha} F_2
\end{equation}
is conserved because of the equation of motion. 
This is nevertheless not sufficient to get a conserved (non positive definite)
product, in fact we must deal with boundary contributions.
In the Euclidean case we consider the surface $S(r_0,r_1)$ in the
upper half plane delimited by two semicircles of ``time'' $r_{0}$ and $r_{1}$
($r_0< r_1$) and by the two segments on the $x$ axis $[r_0,r_1]$ and
$[-r_1,-r_0]$  
then we find\footnote{
Our conventions are
$*d u = -i ~d u$, $*d \bu = i ~d \bu$. 
}
\begin{align}
0= \int_{S(r_0,r_1)} d * j = 
\int_{|u|=r_1} *j  - \int_{|u|=r_0} *j
+ \int_{[r_0,r_1] } *j  + \int_{[-r_1, -r_0] } *j 
.
\end{align}
In order to get a metric independent on time $r$ we need rewriting 
the two integrals along the $x$ axis as a difference of a
function which depends only on the background fields at 
the initial and final times. 
While we know that we can write a definite integral as a difference
of a function evaluated at final and initial times what it is not
certain is that this difference does depend only the final and initial
background fields
since we have  boundary conditions with discontinuities.

The two integrals along the $x$ axis can be expressed 
as a difference of a function which depends only on the background
fields at the evaluation time and  actually vanish
only if we consider solutions $F_q$ which satisfy
the boundary conditions in  eq.s (\ref{global-boundary-upper}), with
$g_t=0$, i.e.
\begin{align}
e^{- i \pi \alpha_t}
 \partial_y \Xz(u,\bu)|_{u=x+ i 0^+ } 
+
e^{ i \pi \alpha_t}
\partial_y \Xbz(u,\bu)|_{u=x+ i 0^+ } 
&= 0
~~~~
x_{t}<x<x_{t-1}
\nonumber\\
e^{-i \pi \alpha_t} \Xz(u,\bu)|_{u=x+ i 0^+ } 
-
e^{ i \pi \alpha_t} \Xbz(u,\bu)|_{u=x+ i 0^+ } 
&= 0
~~~~
x_{t}<x<x_{t-1}
.
\label{quantum-global-boundary-upper}
\end{align}
In the following we call these boundary conditions quantum boundary conditions.
The quantum boundary conditions can be also written as
\begin{align}
(\uno+R_t) \partial_y F_q |_{y=0}=0,~~~~
(\uno-R_t) F_q |_{y=0}=0~~~~
x_t < x < x_{t-1}
\label{qu-global-boundary-upper}
\end{align}
with
\begin{equation}
R_t=R_t^\dagger=R_t^{-1}=
\left(\begin{array}{cc}
& e^{i 2\pi \alpha_{t}} \\ 
 e^{-i 2\pi \alpha_{t}} & 
\end{array}\right) 
.
\end{equation}
Consider then one of the $x$ boundary contributions
\begin{align}
-i \int_{[r_0,r_1] } *j
&=
- \int_{[r_0,r_1] } dx~  j_y
=
\int_{[r_0,r_1] } dx~\left(
\partial_y F_1^\dagger F_2  - F_1^\dagger \partial_y F_2  
\right) \Big|_{y=0}
.
\label{I-r0-r1}
\end{align}
Let us consider the first term $\partial_y F_1^\dagger F_2$. We
can split the integration interval into pieces where the boundary
conditions are constant. Then using the quantum boundary conditions we
have the identity $F_2=\frac{\uno+R_t}{2} F_2$ and therefore we can write
\begin{align}
\partial_y F_1^\dagger F_2  =
\partial_y F_1^\dagger  \frac{\uno+R_t}{2} F_2  =
\left(\frac{\uno+R_t}{2} \partial_y F_1\right)^\dagger F_2  
=0
.
\end{align}
If we would not use the quantum boundary conditions but the original
boundary conditions the second identity in (\ref{qu-global-boundary-upper}) 
would read 
$(\uno-R_t) F |_{y=0}= 2 G_t = - 2 i g_t (e^{2\pi \alpha_{t}}, e^{-i \pi \alpha_{t}} )^T $  
then  the contribution from a piece of the integration interval where 
 the boundary conditions are constant would be
$
-\partial_y F_1^\dagger G_t  + G_t^\dagger \partial_y F_2 
$.
This term can be integrated explicitly using the fact that $F$s split
into a sum of left and right moving pieces and it is non vanishing.
Hence the resulting boundary contribution (\ref{I-r0-r1}) does
depend not only on the fields at $r_0$ and $r_1$ but also on the
fields at the discontinuities between $r_0$ and $r_1$. This means that
the would be product does depend on the history and not only on the
background fields at the time where the metric is computed and hence
there is no time independent product of modes.
We conclude therefore that only for solutions satisfying the quantum
boundary conditions (\ref{qu-global-boundary-upper}) we have the non
positive definite metric (actually Hermitian form)
\begin{align}
(F_1,F_2)
=
(F_2,F_1)^*
&=
\int_{|u|=r}  
\left(
i~F_1^\dagger \stackrel{\leftrightarrow}{\partial_x} F_2 ~dy
-
i~F_1^\dagger \stackrel{\leftrightarrow}{\partial_y} F_2 ~dx
\right)
\nonumber\\
&=
\int_{0; r=const}^\pi  d \theta
~r~
i~\left( F_1^\dagger \stackrel{\leftrightarrow}{\partial_r} F_2 \right)
.
\label{metric_for_modes}
\end{align}
This means that we must proceed as in the path integral approach and
split the string coordinate into a classical and quantum part and then
quantize only the quantum part. 
Explicitly we write
\begin{equation}
X^I(u,\bu)
=
X^I_{cl}(u,\bu;\{x_t,g_t,\alpha_t\})
+
X^I_{q}(u,\bu;\{x_t,\alpha_t\})
\label{Xcl+Xq}
\end{equation}
with $X_{cl}$ satisfying the original boundary conditions
(\ref{global-boundary-upper}) and and $X_{q}$ satisfying the quantum
conditions (\ref{quantum-global-boundary-upper}).
Notice that only with the quantum boundary conditions
(\ref{quantum-global-boundary-upper}) the two dimensional laplacian $\du
\dbu$ is self-adjoint and it is certain to have a
Green function. This is shown in appendix \ref{app:self-adjoint}.

Finally, notice that the previous discussion can be and must be applied
also to branes with magnetic field. In this case the Minkowskian
boundary conditions read $X^{' I}- \cF_{I J s} \dot X^J |_{\sigma=s}=0$
with $s=0,\pi$ and $\cF_{I J s}$ the magnetic fields.
Then the boundary contribution corresponding to  (\ref{I-r0-r1}) is not anymore
zero but evaluates to
\begin{equation}
-i \int_{\tau_0}^{\tau_1} *j 
=
i \int_{\tau_0}^{\tau_1} d \tau~j_\sigma 
=
- F_1^\dagger \cF_{0} F_2|_{\sigma=0,\tau_1}  + F_1^\dagger \cF_{0} F_2 |_{\sigma=0,\tau_0} 
\end{equation}
and therefore induces a product
\begin{equation}
(F_1,F_2)
=
\int_{0}^\pi   i~F_1^\dagger \stackrel{\leftrightarrow}{\partial_\tau}
F_2~d \sigma 
+
i F_1^\dagger \cF_{0} F_2 |_{\sigma=0}
-
i F_1^\dagger \cF_{\pi} F_2 |_{\sigma=\pi}
\end{equation}
which is the ``weird'' metric used in \cite{Abou}.

\subsection{Doubling trick and the metric}
We have established that we must quantize the fluctuation around
the classical solution which satisfy the quantum boundary conditions
(\ref{quantum-global-boundary-upper}) we now look to solutions to equations
of motion with these boundary conditions.
As usual the general solution of eq. (\ref{eom}) is given by 
$X^I(u,\bu)= X^I_L(u)+ X^I_R(\bu)$. Then because of the boundary
conditions we are led to consider the two possible independent sets of
quantum modes
\begin{align}
X_{(c)}(u,\bu)
&=
\vect{\Xz_{q, L}(u)}{\Xbz_{q, R}(\bu)},~~~~
%
%
%
%
%
%
\bX_{(a)}(u,\bu)
=
\vect{\Xz_{q, R}(\bu)}{\Xbz_{q, L}(u)}
%
%
%
\label{X-Xbar}
\end{align}
where $(\Xz_{q, L}(u), \Xbz_{q, R}(\bu))^T$
can be any element of the basis of solutions and
is labeled by a further basis index which is suppressed in this
section. Similarly for $(\Xz_{q, R}(\bu), \Xbz_{q, L}(u))^T$.
After this splitting the couple of quantum boundary conditions in
eq. (\ref{qu-global-boundary-upper}) become simply one condition, one
for each set, explicitly\footnote{
\COMMENTOO{check}
The untwisted case requires
a slightly more general solution because of the $\log|u|$ 
is possible and it is needed for , i.e. we could ask
$\Xz_{q, L}(x+i 0^+)= e^{i 2 \pi \alpha_t} \Xbz_{q, R}(x-i 0^+)+\delta_{L t}$,
$\Xz_{q, R}(x-i 0^+)= e^{i 2 \pi \alpha_t} \Xbz_{q, L}(x+i 0^+)+\delta_{R t}$
but then the non derivative boundary conditions imply that
$\delta_{L t}+\delta_{R t}=0$ and that there is a unique non chiral
solution $X^I_L(u)+ X^I_R(\bu)$.
}
\begin{align}
\Xz_{q, L}(x+i 0^+)= e^{i 2 \pi \alpha_t} \Xbz_{q, R}(x-i 0^+),~~~~
x_t<x<x_{t-1}
\nonumber\\
\Xz_{q, R}(x-i 0^+)= e^{i 2 \pi \alpha_t} \Xbz_{q, L}(x+i 0^+),~~~~
x_t<x<x_{t-1}
\end{align}
since the derivative boundary condition follows from the previous
because $\partial_y X_L|_{y=0^+}= -i \partial_x X_L|_{y=0^+}$ and
$\partial_y X_R|_{y=0^+}= i \partial_x X_R|_{y=0^+}$.
In the case of the classical part the previous equations would have
been stated using the derivatives because there is no obvious way of
splitting the constants $g_t$ into a left and right part. Using
derivatives we miss the information on the constants $g_t$ which has
to be kept adding further conditions. 
Explicitly the boundary conditions for the classical part
can be written as (when all $x_t\ne 0, \infty$)
\begin{align}
\partial \Xz_{cl, L}(x+i 0^+) 
&= 
e^{i 2 \pi \alpha_t} \partial\Xbz_{cl, R}(x-i 0^+),~~~~
x_t<x<x_{t-1}
\nonumber\\
\partial \Xz_{cl, R}(x-i 0^+) 
&= 
e^{i 2 \pi \alpha_t} \partial\Xbz_{cl, L}(x+i 0^+),~~~~
x_t<x<x_{t-1}
\nonumber\\
\Xz_{cl}(x_t,x_t)
&= 
f_t
\end{align}

These $X_{(c)}(u,\bu)$ and $\bX_{(a)}(u,\bu)$ can be combined into two sets of function defined on the whole
complex plane minus the cut $[x_N,x_1]$ using the doubling trick
as\footnote{
Notice that in order to perform the gluing this way we need the $D_1$
to be the last brane on the real positive axis.
}
\begin{align}
\cX(z)
&=
\left\{
\begin{array}{cc}
   \Xz_{q, L}(u) 
  & z=u\mbox{ with }{Im~} z >0\mbox{ or } z\in\R-[x_N,x_1] 
  \\
  e^{i 2\pi \alpha_1} \Xbz_{q, R}(\bar u) 
  & z=\bar u\mbox{ with }{Im~} z <0\mbox{ or } z\in\R-[x_N,x_1] 
\end{array}
\right.
\nonumber\\
%
\cbX(z)
&=
\left\{
\begin{array}{cc}
  \Xbz_{q, L}(u) 
  & 
  z= u\mbox{ with }{Im~} z >0\mbox{ or } z\in\R-[x_N,x_1] 
  \\
  e^{-i 2\pi \alpha_1} \Xz_{q, R}(\bar u) 
  & 
  z=\bar u\mbox{ with }{Im~} z <0\mbox{ or } z\in\R-[x_N,x_1] 
\end{array}
\right.
\label{calX-calXbar}
\end{align}
with $\cbX(z)\ne (\cX(z))^*$ and very simple boundary conditions
\begin{align}
\left\{\begin{array}{c c}
\cX(x+i0^+)=\cX(x+i0^-) & x<x_N \mbox{ and } x>x_1
\\
\cX(x+i0^+)=e^{i 2\pi (\alpha_t -\alpha_1)} \cX(x+i0^-) & x_t <x <
x_{t-1} \mbox{ for } t=2\dots N
\end{array}\right.,
\nonumber\\
\left\{\begin{array}{c c}
\cbX(x+i0^+)=\cbX(x+i0^-) & x<x_N \mbox{ and } x>x_1
\\
\cbX(x+i0^+)=e^{-i 2\pi (\alpha_t -\alpha_1)} \cbX(x+i0^-) & x_t <x <
x_{t-1} \mbox{ for } t=2\dots N
\end{array}\right.
.
\end{align}

Obviously the  gluing can be performed in different ways,
i.e. we can  glue along $D_\bt$ brane instead of $D_1$ as
\begin{align}
\cX^{(\bt)}(z)
&=
\left\{
\begin{array}{cc}
   \Xz_{q, L}(u) 
  & z=u\mbox{ with }{Im~} z >0\mbox{ or } 
   z\in\R-(-\infty,x_\bt]-[x_{\bt+1}, \infty) 
  \\
  e^{i 2\pi \alpha_\bt} \Xbz_{q, R}(\bar u) 
  & z=\bar u\mbox{ with }{Im~} z <0\mbox{ or } 
   z\in\R-(-\infty,x_\bt]-[x_{\bt+1}, \infty) 
\end{array}
\right.
\nonumber\\
%
\cbX^{(\bt)}(z)
&=
\left\{
\begin{array}{cc}
  \Xbz_{q, L}(u) 
  & 
  z= u\mbox{ with }{Im~} z >0\mbox{ or } 
   z\in\R-(-\infty,x_\bt]-[x_{\bt+1}, \infty) 
  \\
  e^{-i 2\pi \alpha_\bt} \Xz_{q, R}(\bar u) 
  & 
  z=\bar u\mbox{ with }{Im~} z <0\mbox{ or }
   z\in\R-(-\infty,x_\bt]-[x_{\bt+1}, \infty) 
\end{array}
\right.
\label{calX-calXbar-generic}
\end{align}
again with $\cbX^{(\bt)}(z)\ne (\cX^{(\bt)}(z))^*$ and  boundary conditions
\begin{align}
\left\{\begin{array}{c c}
\cX(x+i0^+)=\cX(x+i0^-) & 
x_\bt<x<x_{\bt-1}
\\
\cX(x+i0^+)=e^{i 2\pi (\alpha_t -\alpha_\bt)} \cX(x+i0^-) 
& x_t <x < x_{t-1} \mbox{ for } t=1\dots N,~~t\ne\bt
\end{array}\right.,
\nonumber\\
\left\{\begin{array}{c c}
\cbX(x+i0^+)=\cbX(x+i0^-) & x_\bt<x<x_{\bt-1}
\\
\cbX(x+i0^+)=e^{-i 2\pi (\alpha_t -\alpha_1)} \cbX(x+i0^-) 
& x_t <x < x_{t-1} \mbox{ for } t=1\dots N~~t\ne\bt
\end{array}\right.
.
\end{align}
In the following we use always the gluing along $D_1$ if not otherwise stated.

The metric (\ref{metric_for_modes}) can then be calculated for any
pairs of these functions using the doubled formalism
as (for more details see appendix \ref{app:details_metric})
\begin{align}
(X_{(c) 1}, X_{(c) 2})
&=
(\bX_{(a) 1}, \bX_{(a) 2})
=0
%
\label{cXcX}
\\
(X_{(c)}, \bX_{(a)})
&=
2 e^{i 2\pi \alpha_1}
\oint_{z= r_0 \exp{i \theta}; \theta\in[-\pi,\pi]}
d z~ (\cX(\bz))^* \frac{d \cbX(z) }{d z}
\nonumber\\
&=
2 e^{i 2\pi \alpha_1}
\int_{-\pi}^{\pi} d \theta~ (\cX(r_0 e^{-i \theta} ))^* 
\frac{d \cbX( r_0 e^{i \theta} ) }{d \theta}
,
\label{cXcbX}
\end{align}
where the last equation is meaningful since the product 
$(\cX(r_0 e^{-i \theta} ))^* \frac{d \cbX( r_0 e^{i \theta} ) }{ d \theta}$
is continuous for $\theta=0$ despite the fact its factors are not.
A direct computation similar to the one done to get the previous
equation (\ref{cXcbX}) gives 
\begin{equation}
(\bX_{(a)}, X_{(c)})
=
2 e^{-i 2\pi \alpha_1}
\int_{-\pi}^{\pi} d \theta~ (\cbX(r_0 e^{-i \theta} ))^* 
\frac{d \cX( r_0 e^{i \theta} ) }{d \theta}
\end{equation}
which is obviously compatible with the Hermitian property of the form and the
product $(X_{(c)}, \bX_{(a)})$ in eq. (\ref{cXcbX}).

\subsection{Radial canonical quantization}
We want now quantize the Euclidean string action (\ref{S_E}).
In order to do so we split the string field into its classical and
quantum part (\ref{Xcl+Xq}) so that the action becomes
\begin{equation}
S_E
=
S_{E, cl}
+
\frac{1}{4\pi \alpha'} \int_H d \theta d r~r~ 
(\partial_r \Xz_q \partial_r \Xbz_q 
+ r^{-2} \partial_\theta \Xz_q  \partial_\theta \Xbz_q )
.
\label{S_Eq}
\end{equation}
We take $\tau_E=\ln r$ to be the time but we write all expressions as
functions of $r$, this means that the Hamiltonian rescales $r=|u|=e^{\tau_E}$ and
not that it shifts $r$.
The canonical momentum is then given by 
\begin{align}
P_q
= \vect {P_{q\, \bz}}{P_{q\, z}}
= \frac{r}{2 \pi \alpha'} \vect { \partial_r \Xz_q } { \partial_r
  \Xbz_q }
,
\end{align}
and the Euclidean Hamiltonian is by definition  
\begin{equation}
H= 
 \int_0^\pi d \theta~
\left(
{\pi \alpha'} {P_{q\, \bz}}{P_{q\, z}}
+ 
\frac{1}{4\pi \alpha'}
\partial_\theta \Xz_q  \partial_\theta \Xbz_q 
\right)
.
\label{Eucl-Ham}
\end{equation}
From the canonical commutation relation
\begin{equation}
[ X^I_q(\theta), P_{q\, J}(\theta') ] = i \delta^I_J \delta(\theta-\theta')
\end{equation}
with $\delta(\theta-\theta')$ the delta function with the appropriate
boundary conditions, we get
\begin{equation}
[H, X_q(\theta)]= -i  r \partial_r X_q(\theta),~~~~
[H, P_q(\theta)]= -i r \partial_r P_q(\theta),~~~~
\end{equation}

In order to write the Hamiltonian using the creation and annihilation
operators  we need a way to extract them from the quantum field $X_q$
by mean of a product of the quantum fluctuation $X_q$ 
with an appropriate solution $F$. This product can be written as
\begin{equation}
(F,X_q)
=
i \int_0^\pi d \theta \left( 
2 \pi \alpha' F^\dagger P_q -r \partial_r F^\dagger X_q \right)
.
\end{equation}
In particular the commutation relation of two such products is given by
\begin{align}
[(F_1, X_q), (F_2, X_q) ]
= 
-2 \pi \alpha' ~i~
(F_1, \sigma_1 F_2^*)
,
\end{align} 
where $\sigma_1$ is the Pauli matrix.
When choosing the two solutions $F_{1,2}$ to be 
any of the basis elements the previous commutation relation become 
\begin{align}
[(X_{(c) n}, X_q), (X_{(c) m}, X_q) ]
&= 
[(\bX_{(a) n}, X_q), (\bX_{(a) m}, X_q) ]
=0
\nonumber\\
[(X_{(c) n}, X_q), (\bX_{(a) m}, X_q) ]
&=
-2 \pi \alpha' ~i~
(X_{(c) n}, \sigma_1 \bX_{(a) m}^*)
\label{comm-rel-v0}
\end{align} 

$X_{(c) n}$ and $\bX_{(c) n}$ are a basis for the quantum modes.
In deriving the first  equation in the first line  we used the fact that
any $\sigma_1 X_{(c) n}^*$ has the same boundary conditions as any of
$X_{(c) m}$ and hence they can be expanded on the $X_{(c) m}$ basis.
Then we can use eq. (\ref{cXcX}) to set to zero the commutation
relations involving two $X_{(c)}$ modes.
Similarly for $\bX_{(a) n}$.

We can now simplify the previous equations (\ref{comm-rel-v0}) 
if we notice that the function $\bar \chi_{\bullet n}(z)$ 
associated with $ \sigma_1 \bX_{(c) n}^*$ by the doubling trick 
can be rewritten as
\begin{align}
\bar \chi_{\bullet n}(z)= e^{-i 2\pi \alpha_1} \left[ \bar\chi_n(\bz) \right]^*
\end{align}
and similarly for 
$\chi_{\bullet n}(z)= e^{i 2\pi \alpha_1} \left[\chi_n(\bz) \right]^*$ 
which is associated with $ \sigma_1 X_{(c) n}^*$.
Using 
these relations, the hermiticity property of the product
and the explicit expression in eq. (\ref{cXcbX}) for the product in
terms of the doubled mode functions we can then
write the non vanishing commutation relations for the basis elements
(\ref{comm-rel-v0}) as
\begin{align}
[(\bX_{(a) n}, X_q), (X_{(c) m}, X_q) ]
&=
-4 \pi \alpha' ~i~
\left[
\oint_{|z|=r_0}
d z~ \cbX_{m}(z) \frac{d \cX_n(z) }{d z}
\right]^*
.
\label{comm-rel-v1}
\end{align}

We now expand the quantum fluctuation as\footnote{
Notice that generically neither $X$ nor its derivatives $\partial_u X$
are conformal fields since they are not well behaved under time
evolution $u\rightarrow \lambda u$.}
\begin{align}
X_q(u,\bu)
&= 
\sum_{n\in\Z} \left( x_n X_{(c) n} + \bar x_n \bX_{(a) n} \right)
.
\end{align}
We suppose that the (doubling of) quantum modes satisfy 
a reality condition like
\begin{equation}
\left( \chi_n(\bz) \right)^*= e^{i \beta} \chi_n(z),~~~~
\left( \bar\chi_n(\bz) \right)^*= e^{i \hat\beta} \bar\chi_n(z)
,~~~~ \beta,\hat\beta\in\R
\end{equation}
and a normalization condition
\begin{equation}
(X_{(c) n}, \sigma_1 \bX^*_{(a) m}) = -N_n \delta_{n+m,s}
\end{equation}
with $s$ an integer and $N_n^*=- N_n e^{-i \beta -i \hat \beta}$.
It follows that the creation and annihilation operators can be
obtained as
\begin{align}
x_n = \frac{e^{i 2\pi \alpha_1+ i \beta} }{N_{s-n}} (\bX_{(a) s-n}, X_q),~~~~
\bx_n = -\frac{e^{i 2\pi \alpha_1+ i \hat \beta} }{N_{s-n}} (X_{(c) s-n}, X_q)
,
\end{align}
and that they satisfy the commutation relations
\begin{align}
[x_n, \bx_m]
=
i 2 \pi \alpha' \frac{ e^{i \beta + i \hat\beta} }{N_{s-n}} \delta_{n+m,s}
.
\end{align}

On general ground the Euclidean Hamiltonian defined in eq. (\ref{Eucl-Ham})
is generically {\sl not} diagonal in the mode operators since the
Lagrangian is time dependent
or that is the same in states are not equal to out states and hence
they cannot be eigenstates of the Hamiltonian.
Another way to understand this is to notice that the Hamiltonian
rescales $u$ but the generic mode function is not an homogeneous
function of $u$.
Explicitly the Hamiltonian can be expanded in modes as
\begin{equation}
H= \sum_{n,m\in\Z}h_{n m} x_n \bx_m,~~~
 h_{n m}=\frac{1}{2\pi \alpha'} e^{i 2\pi \alpha_1} 
(\bX_{(a) n}, r \partial_r X_{(c) m}) 
\end{equation}
where $h_{n m}$ is constant despite the Lagrangian is time dependent
because of the boundary conditions. This happens because
 $r \partial_r X_{(c) m}$ are also  solutions of the e.o.m. and
therefore satisfy the same ``selection rules'' as $X_{(c) m}$.

$H$ is not diagonal in modes even if it is self-adjoint w.r.t. the
usual $L^2$ metric because it is self-adjoint for any time $\tau$ but
it does depend on time.

Finally, notice that we have not written any normal ordering since its
definition depends on the vacuum and we have not specified
any. Neither we will do it since we will use an overlap approach which
uses different Hamiltonian for different worldsheet times. See sections
\ref{sect:normal_modes_overlap} and \ref{Green_function_in_out_vacua}.

\subsection{The $N=0$ case: the usual untwisted string}
Since the previous product for the modes is different from the normal
one it is of interest to see how it works in the usual 
and simplest case with ND boundary conditions. 
It is also worth to check that we get the same commutation
relations as in the quantization with the normal product.

We consider a single $D1$ brane $D_t$ in $\R^2$.
In this case the boundary conditions are simply for all $x\in\R$
\begin{align}
e^{- i \pi \alpha_t}
 \partial_y \Xz(u,\bu)|_{u=x+ i 0^+ } 
&+
e^{ i \pi \alpha_t}
\partial_y \Xbz(u,\bu)|_{u=x+ i 0^+ } 
= 0
,
\nonumber\\
e^{-i \pi \alpha_t} \Xz(u,\bu)|_{u=x+ i 0^+ } 
&-
e^{ i \pi \alpha_t} \Xbz(u,\bu)|_{u=x+ i 0^+ } 
= 
2 i g_t
.
\end{align}
The classical solution is simply given by
\begin{align}
X_{cl}(u,\bu; \{g_t,\alpha_t\})=  e^{i \pi \alpha_t}\, i\, g_t  
\vect{ 1 } { - e^{-i 2\pi \alpha_t}  }
.
\end{align}
The two sets of modes in eq.s (\ref{X-Xbar}) which obey the quantum boundary
conditions are
\begin{align}
X_{(c) n}(u,\bu; \{\alpha_t\})
=
\vect { \frac{u^{-n}}{n} } { e^{-i 2\pi \alpha_t} \frac{\bu^{-n}}{n} },~~~
\bX_{(a) n}(u,\bu; \{\alpha_t\})
=
\vect{ \frac{\bu^{-n}}{n} } { e^{-i 2\pi \alpha_t} \frac{u^{-n}}{n} },
~~~
n\ne0
\end{align}
and 
\begin{equation}
\hat X_{ 0}(u, \bu; \{\alpha_t\}) 
=
\vect{ \log |u| } { e^{-i 2\pi \alpha_t} \log |u| },~~~
X_{*}(\{\alpha_t\}) 
=
\vect{ 1 } { e^{-i 2\pi \alpha_t}  }
.
\end{equation}
Notice however that $\hat X_{ 0}$ 
is different from all the others elements since its components are
neither holomorphic nor antiholomorphic since neither
 the holomorphic  nor the antiholomorphic parts satisfy the boundary
conditions separately.

Furthermore $X_*$ is not proportional to the classical solution
$X_{cl}$, as it could at first glance look  because of the sign of the
second component.  

Using the doubling trick
the previous modes can be combined into the following functions
defined in the whole complex plane 
\begin{align}
\cX_n(z) = \frac{z^{-n}}{n},~~~
\cbX_n(z) = e^{-i 2\pi \alpha_t} \frac{z^{-n}}{n},~~~
\cX_0(z,\bz)= \log |z|,~~~
\cX_*= 1
.
\label{N=0-basis}
\end{align}
Using these functions and eq.s (\ref{cXcX}) and (\ref{cXcbX}) we can compute
the non vanishing products of these elements
\begin{align}
(X_{(c) n},\bX_{(a) m} )
& 
= \left[ (\bX_{(a) m},X_{(c) n} ) \right]^*
= -\frac{4\pi i}{n} \delta_{n+m,0}
\nonumber\\
(X_{*}, \hat X_{ 0} )
&
=\left[ ( \hat X_{ 0}, X_{*} ) \right]^*
= -2 \pi i
,
\end{align}
where all products involving $\hat X_0$ cannot be computed using
eq. (\ref{cXcbX}).
In fact this equation is derived under the hypothesis that the two
functions can be assembled into (anti)holomorphic doubled functions
defined on the complex plane.
Therefore these products must be
computed from the original definition given in eq. (\ref{metric_for_modes}).

We can now expand the quantum fluctuations as
\begin{align}
X_q(u,\bu)
&= 
x_0 X_{*} + \hat x_0  \hat X_{ 0}
+\sum_{n\ne0} \left( x_n X_{(c) n} + \bar x_n \bX_{(a) n} \right)
\nonumber\\
&=
\vect{
x_0 + \hat x_0 \ln |u|
+\sum_{n\ne0} \left( x_n \frac{u^{-n}}{n} + \bar x_n \frac{\bu^{-n}}{n} \right)
}{
e^{-i 2\pi \alpha_t}\left(
x_0 + \hat x_0 \ln |u|
+\sum_{n\ne0} \left( x_n \frac{\bu^{-n}}{n} + \bar x_n \frac{u^{-n}}{n} \right)
\right)
}
.
\end{align}
The operators can be extracted from the previous expansion as
\begin{align}
x_0 &= \frac{1}{2 \pi  i} (\hat X_0, X_q) 
\nonumber\\
\hat x_0 &= -\frac{1}{2\pi i } (X_*, X_q) 
\nonumber\\
x_n &= \frac{-n}{ 4\pi i} (\bX_{(a) -n}, X_q) 
\nonumber\\
\bar x_n &= \frac{n}{ 4\pi i} (X_{(c) -n}, X_q) 
,
\end{align}
and then we can compute the non vanishing commutation relations as
\begin{align}
[ x_{n}, \bar x_{m} ]
&=
 e^{i 2\pi \alpha_t } \frac{\alpha'}{2} m \delta_{n+m,0}
\nonumber\\
[ x_{0}, \hat x_{0} ]
&=
\alpha'  e^{i 2\pi \alpha_t }
.
\end{align}
These are exactly the usual commutation relations and expansion 
once we identify ($n>0$)
\begin{align}
x_n= i \frac{\sqrt{2\alpha'}}{2}  e^{i\pi \alpha_t } \bar \alpha_n,~~~
&
x_{-n}= i \frac{\sqrt{2\alpha'}}{2}  e^{i\pi \alpha_t } \alpha_{-n},~~~
\nonumber\\
\bar x_n= i \frac{\sqrt{2\alpha'}}{2}  e^{i\pi \alpha_t } \alpha_n,~~~
&
\bar x_{-n}= i \frac{\sqrt{2\alpha'}}{2}  e^{i\pi \alpha_t } \bar\alpha_{-n},~~~
\nonumber\\
x_0 =  e^{i \pi \alpha_t } \frac{x^{\hat 1}}{\sqrt{2}},~~~
&
\hat x_0 =  e^{i \pi \alpha_t } \frac{-i 2\alpha' p^{\hat 1}}{\sqrt{2}}
.
\end{align}
As usual the vacuum is defined as
\begin{align}
p^{\hat 1} |0\rangle
=
\alpha_n |0\rangle
=
\bar \alpha_n |0\rangle
=0~~~~
n>0
.
\end{align}
We can then compute the untwisted Green functions
\begin{align}
G^{z z}_{U_t}(u,\bu; v,\bv; \alpha_t)
&= 
[X^{(+)}(u,\bu), X^{(-)}(v,\bv)]
=\left( -i\oh\sqrt{2\alpha'} e^{i \pi \alpha_t} \right)^2 \ln |u-\bv|^2
\nonumber\\
G^{\bz \bz}_{U_t}(u,\bu; v,\bv; \alpha_t)
&= 
[\bar X^{(+)}(u,\bu), \bar X^{(-)}(v,\bv)]
= \left( -i\oh\sqrt{2\alpha'} e^{-i \pi \alpha_t} \right)^2 \ln |u-\bv|^2
\nonumber\\
G^{z \bz}_{U_t}(u,\bu; v,\bv; \alpha_t)
&= 
[X^{(+)}(u,\bu), \bar X^{(-)}(v,\bv)]
=\left( -i\oh\sqrt{2\alpha'} \right)^2 \ln |u-v|^2
.
\label{untwisted-green-function}
\end{align}
Notice that $G^{z \bz}_{U}$ does not feel whether the brane is rotated
while both $G^{z z}_{U}$ and $G^{\bz \bz}_{U}$ do because of the
phases.

Finally, we consider two parallel branes non
overlapping then the non derivative boundary conditions become
\begin{align}
e^{-i \pi \alpha_t} \Xz(u,\bu)|_{u=x+ i 0^+ } 
&-
e^{ i \pi \alpha_t} \Xbz(u,\bu)|_{u=x+ i 0^+ } 
= 
\sqrt{2} i g_{t+1},~~~~
x<0
\nonumber\\
e^{-i \pi \alpha_t} \Xz(u,\bu)|_{u=x+ i 0^+ } 
&-
e^{ i \pi \alpha_t} \Xbz(u,\bu)|_{u=x+ i 0^+ } 
= 
\sqrt{2} i g_t,~~~~
x>0
\end{align}
while the derivative one is left unchanged.
The classical solution is then
\begin{align}
X_{cl}(u,\bu; \{g_t,\alpha_t\})
=  
e^{i \pi \alpha}\, i\, g_t  \vect{ 1 } { - e^{-i 2\pi \alpha_t}  }
+
e^{i \pi \alpha} \frac{ g_{t+1}- g_t}{ \pi}  
\vect{ \oh \ln \frac{u}{\bu} } {  e^{-i 2\pi \alpha_t}\,  \oh \ln \frac{\bu}{u} }
\end{align}
while the quantum fluctuations remain unchanged.
Notice however that the classical solution has infinite action because
of the finite and constant energy density of the stretched string.

\subsection{A $N=2$ case: the usual twisted string}
We now consider the quantization of two $D1$ branes at
angles, $D_t$ and $D_{t+1}$.
This is the usual setup where the string describes a twisted in and out state.
This means that the twist fields are located at $x=0$ and $x=\infty$
so that the boundary conditions read
\begin{align}
&\mycases{
e^{-i \pi \alpha_{t+1}} \partial_y \Xz(u,\bu)|_{u=x+ i 0^+ } 
+
e^{ i \pi \alpha_{t+1}} \partial_y \Xbz(u,\bu)|_{u=x+ i 0^+ } 
= 0
}{
e^{-i \pi \alpha_{t+1}} \Xz(u,\bu)|_{u=x+ i 0^+ } 
-
e^{ i \pi \alpha_{t+1}} \Xbz(u,\bu)|_{u=x+ i 0^+ } 
= 
2 i g_{t+1}
},~~~~
x<0
\nonumber\\
&\mycases{
e^{-i \pi \alpha_t} \partial_y \Xz(u,\bu)|_{u=x+ i 0^+ } 
+
e^{ i \pi \alpha_t} \partial_y \Xbz(u,\bu)|_{u=x+ i 0^+ } 
= 0
}{
e^{-i \pi \alpha_t} \Xz(u,\bu)|_{u=x+ i 0^+ } 
-
e^{ i \pi \alpha_t} \Xbz(u,\bu)|_{u=x+ i 0^+ } 
= 
2 i g_t
},~~~~
x>0
.
\end{align}
The classical solution is then simply given by the constant
\begin{align}
X_{cl}(u,\bu; \{g_t, \alpha_t; g_{t+1}, \alpha_{t+1}\})
=\vect{f_t}{f_t^*},~~~~
f_t= \frac{e^{i \pi \alpha_{t+1}} g_t - e^{i \pi \alpha_{t}} g_{t+1}
}{ 
\sin \pi ( \alpha_{t+1} - \alpha_{t} )
}
\label{N=2-X-cl}
\end{align}
which is a special case of eq. (\ref{f-intersection}).
The quantum fluctuations can be expanded on
\begin{align}
X_{(c) n}(u, \bu; \{\alpha_t,\alpha_{t+1}\})
=
\vect { \frac{u^{-n-\bep_t}}{n+\bep_t} } 
{ e^{-i 2\pi \alpha_t} \frac{\bu^{-n-\bep_t}}{n+\bep_t} },~~~
\bX_{(a) n}(u, \bu; \{\alpha_t,\alpha_{t+1}\})
=
\vect{ \frac{\bu^{-n-\epsilon_t}}{n+\epsilon_t} } 
{ e^{-i 2\pi \alpha_t} \frac{u^{-n-\epsilon_t}}{n+\epsilon_t} }.
\end{align}
These can be combined into the function defined on the whole complex
plane  minus the real negative axis as follows from
eq. (\ref{calX-calXbar})\footnote{
We choose $D_t$ to be on the real positive axis in order to be able to
apply this general formula, in particular this means that the cut is on
the negative real axis and $-\pi<arg(z)<\pi$.}
\begin{align}
\cX_n(z)= \frac{z^{-n-\bep_t}}{n+\bep_t},~~~
\cbX_n(z)= e^{-i 2\pi \alpha_t} \frac{z^{-n-\epsilon_t}}{n+\epsilon_t},~~~
z\in\C-\R^-
\label{cX-usual-N=2-twisted}
\end{align}
with $\epsilon_t= \alpha_{t+1}-\alpha_t+ \theta(\alpha_t-\alpha_{t+1})$ and
$\bep_t=1-\epsilon_t$ so that $0<\epsilon_t,\bep_t<1$.

The non vanishing products of these elements are
\begin{align}
(X_{(c) n},\bX_{(a) m} )
& 
= \left[ (\bX_{(a) m},X_{(c) n} ) \right]^*
=- \frac{4\pi i}{n+\bep_t} \delta_{n+m+1,0}
.
\end{align}
As in the previous case the quantum fluctuations can be expanded as
\begin{align}
X_q(u,\bu)
&= 
\sum_{n\in\Z} \left( x_n X_{(c) n}(u,\bu) + \bar x_n \bX_{(a) n}(u,\bu) \right)
\nonumber\\
&=
\vect{
\sum_{n\in\Z} \left( x_n \frac{u^{-n-\bep_t}}{n+\bep_t} 
+ \bar x_n \frac{\bu^{-n-\epsilon_t}}{n+\epsilon_t} \right)
}{
e^{-i 2\pi \alpha_t}
\sum_{n\in\Z} \left( x_n \frac{\bu^{-n-\bep_t}}{n+\bep_t} 
+ \bar x_n \frac{u^{-n-\epsilon_t}}{n+\epsilon_t} \right)
}
.
\end{align}
The coefficients can then be extracted as
\begin{align}
x_{n}
&=
\frac{n+\bep_t}{4 \pi i} (\bX_{(a) -n-1}, X_q)
\nonumber\\
\bar x_{n}
&=
\frac{n+\epsilon_t}{4 \pi i} (X_{(c) -n-1}, X_q)
\end{align}
and their non vanishing commutation relations are
\begin{equation}
[\bar x_n, x_m]= \frac{\alpha'}{2} e^{i 2\pi \alpha_t} (m+\bep_t) \delta_{m+n+1,0}
.
\end{equation}
With the identification ($n\ge0$)
\begin{align}
x_n= i \frac{\sqrt{2\alpha'}}{2}  e^{i\pi \alpha_t } \bar \alpha_{n+\bep_t},~~~
&
x_{-(n+1)}= i \frac{\sqrt{2\alpha'}}{2}  e^{i\pi \alpha_t } \alpha_{n+\epsilon_t}^\dagger,~~~
\nonumber\\
\bar x_n= i \frac{\sqrt{2\alpha'}}{2}  e^{i\pi \alpha_t } \alpha_{n+\epsilon_t},~~~
&
\bar x_{-(n+1)}= i \frac{\sqrt{2\alpha'}}{2}  e^{i\pi \alpha_t }
\bar\alpha_{n+\bep_t}^\dagger
,
\end{align}
we recover the usual commutation relations
\begin{equation}
[ \alpha_{ n+\epsilon_t}, \alpha_{ m+\epsilon_t}^\dagger]
=( n+\epsilon_t) \delta_{m,n}
,~~~~
[ \bar\alpha_{ n+\bar\epsilon_t}, \bar\alpha_{ m+\bar\epsilon_t}^\dagger]
=( n+\bar\epsilon_t) \delta_{m,n}
.
\end{equation}

Finally, we can write the usual expansion for the quantum fluctuations
as 
\begin{align}
X_q(u,\bu)
=
&
i\oh \sqrt{2\alpha'}
\sum_{n=0}^\infty
\vect{
e^{i \pi \alpha_t}
\left[ 
\frac{\bar \alpha_{n+\bar\epsilon_t}}{ n+\bar\epsilon_t} u^{-(n+\bar\epsilon_t)}
-
\frac{\alpha_{n+\epsilon_t}^\dagger}{ n+\epsilon_t} u^{n+\epsilon_t}
\right]
}
{
e^{-i \pi \alpha_t}
\left[
-
\frac{\bar \alpha_{n+\bar\epsilon_t}^\dagger}{ n+\bar\epsilon_t} u^{n+\bar\epsilon_t}
+
\frac{\alpha_{n+\epsilon_t}}{ n+\epsilon_t} u^{-(n+\epsilon_t)}
\right]
}
\nonumber\\
&+
i \oh \sqrt{2\alpha'}
\sum_{n=0}^\infty 
\vect{
e^{i \pi \alpha_t}
\left[
-
\frac{\bar \alpha_{n+\bar\epsilon_t}^\dagger}{ n+\bar\epsilon_t} \bu^{n+\bar\epsilon_t}
+
\frac{\alpha_{n+\epsilon_t}}{ n+\epsilon_t} \bu^{-(n+\epsilon_t)}
\right]
}
{
e^{-i \pi \alpha_t}
\left[ 
\frac{\bar \alpha_{n+\bar\epsilon_t}}{ n+\bar\epsilon_t} \bu^{-(n+\bar\epsilon_t)}
-
\frac{\alpha_{n+\epsilon_t}^\dagger}{ n+\epsilon_t} \bu^{n+\epsilon_t}
\right]
}
.
\label{loc-exp-X}
\end{align}

The vacuum is defined in the usual way by
\begin{equation}
\alpha_{ n+\epsilon_t} |T_t\rangle =\bar \alpha_{ n+\bar\epsilon_t }|T_t\rangle =0
~~~~ n\ge 0
.
\label{twisted-vacuum}
\end{equation}
We can then compute the $N=2$ twisted Green functions
\begin{align}
%
%
G^{z z}_{N=2, T}& (u,\bu; v,\bv; \PaMoSpNDue)
= 
[X_q^{z (+)}(u,\bu), X_q^{z (-)}(v,\bv)]
\nonumber\\
&=
-\left( -i\oh\sqrt{2\alpha'} e^{i \pi \alpha_t} \right)^2
\left[
\frac{1}{\epsilon_t} \left(\frac{v }{ \bu}\right)^{\epsilon_t}
{}_2F_1(1,\epsilon_t; 1+\epsilon_t; \frac{v }{ \bu})
+
\frac{1}{\bep_t} \left(\frac{\bv }{ u}\right)^{\bep_t}
{}_2F_1(1,\bep_t ;1+\bep_t; \frac{\bv }{ u})
\right]
\nonumber\\
%
G^{\bz \bz}_{N=2, T}& (u,\bu; v,\bv; \PaMoSpNDue)
= 
[X_q^{\bz (+)}(u,\bu), X_q^{\bz (-)}(v,\bv)]
\nonumber\\
&=
-\left( -i\oh\sqrt{2\alpha'} e^{-i \pi \alpha_t} \right)^2
\left[
\frac{1}{\epsilon_t} \left(\frac{\bv }{ u}\right)^{\epsilon_t}
{}_2F_1(1,\epsilon_t ;1+\epsilon_t; \frac{\bv }{ u})
+
\frac{1}{\bep_t} \left(\frac{v }{ \bu}\right)^{\bep_t}
{}_2F_1(1,\bep_t; 1+\bep_t; \frac{v }{ \bu})
\right]
\nonumber\\
G^{z \bz}_{N=2, T}& (u,\bu; v,\bv; \PaMoSpNDue)
= 
[ X_q^{z (+)}(u,\bu), X_q^{\bz (-)}(v,\bv)]
\nonumber\\
&=
G^{\bz z}_{N=2}(v,\bv; u,\bu; \PaMoSpNDue)
=
G^{\bz z}_{N=2}(u,\bu; v,\bv; \PaMoSpNDue)
\nonumber\\
&=
-\left( -i\oh\sqrt{2\alpha'} \right)^2
\left[
\frac{1}{\epsilon_t} \left(\frac{\bv }{ \bu}\right)^{\epsilon_t}
{}_2F_1(1,\epsilon_t; 1+\epsilon_t; \frac{\bv }{ \bu})
+
\frac{1}{\bep_t} \left(\frac{v }{ u}\right)^{\bep_t}
{}_2F_1(1,\bep_t; 1+\bep_t; \frac{v }{ u})
\right]
,
\label{Green-N=2}
\end{align}
where we have used
\begin{equation}
{}_2F_1(1,\epsilon;1+\epsilon; x)
=
\sum_{n=0}^\infty \frac{\epsilon}{n+\epsilon} x^n
~~~~
|x|<1
\end{equation}
as follows from the general expression for the hypergeometric function
${}_2F_1(a, b; c; x)= \sum_{n=0}^\infty \frac{ (a)_n (b)_n}{ n! (c)_n}x^n$ 
with $(a)_n= \Gamma(a+n) /\Gamma(a)$ the Pochhammer symbol.
They have the following symmetry properties
\begin{align}
G^{I J}_{N=2, T}(u,\bu; v,\bv; \PaMoSpNDue)
&=
G^{J I}_{N=2, T}( v,\bv; u,\bu; \PaMoSpNDue)
\nonumber\\
&=
G^{I J}_{N=2, T}( v,\bv; u,\bu; \PaMoSpNDue)
\label{Green-N=2-symmetry}
\end{align}
which follow from the hypergeometric transformation properties,
in particular
${}_2 F_1(1, \epsilon; 1+\epsilon; x) =  
\frac{\epsilon }{\bar \epsilon x}
{}_2F_1(1, \bar\epsilon; 1+\bar\epsilon; 1/x)$ implies the last equality.

\subsection{A $N=3$ case: in and out twisted strings}
\label{sect:N=3-in-out-twisted}
In this section we want to exam the next simplest example which
corresponds to the case of $N=3$ twists, one located in the origin,
one at infinity and the third in an arbitrary point of the positive
real axis.
This example shows the main issues to be understood and solved.
When we look after the classical solution with finite action 
it is naturale to consider a basis of the derivatives which can be
used to compute  the classical solution like  $\partial_z \chi_{cl}(z)$.
Nevertheless when we try to use their 
integrated expressions $\int d z \partial_z \chi_{cl}(z)$ 
as a basis 
for the quantum fluctuations we realize immediately that  they
do not satisfy all the boundary conditions.
We are therefore forced to consider a combination of them but in doing
so it seems naively that we cannot find all possible asymptotics
behaviors for $u \rightarrow 0$.
The apparent solution of this problem seems to be very simple.
It amounts to  consider from start a basis $\chi_q(z)$ 
and not a basis for derivatives.
This basis is however not apt to easily find the classical
solution since the configuration with finite action is the combination
of infinite basis elements.
Moreover the simplest basis is not orthogonal and therefore we must
orthogonalize it. In doing so combinations of infinite basis elements
must be considered. 
It happens then that some new basis elements have a finite convergence
radius and must be analytically continued. In performing the
analytic continuation we find that these new basis elements do not
satisfy anymore the original and required boundary conditions.
Hence one of the new basis element must be used to make further combinations
which respect the boundary conditions and in order to preserve
orthogonality it must be dropped as independent basis element.
In this way we find again the original problem, i.e. we cannot find
all possible asymptotics behaviors for $u \rightarrow 0$.
Since we are not yet in stand of understanding completely this issue
we use the classical overlap approach to derive the correlators. 

Let us now explain in more details the issues discussed above.

\begin{figure}[hbt]
\begin{center}
\def\svgwidth{300px}
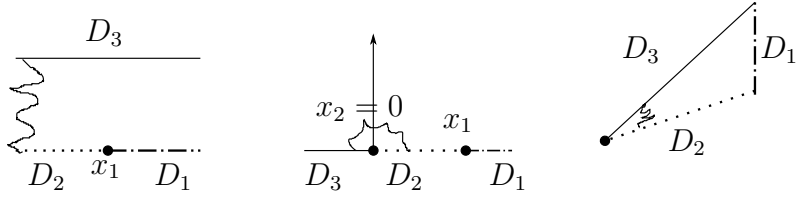
\end{center}
\vskip -0.5cm
\caption{Three different pictures of the setup with incoming and
outgoing twisted strings with a discontinuity on the boundary at
$x_1$.
In the pictures an incoming string between $D_2$ and $D_3$ is pictured.
}
\label{fig:N3_in_out_twisted}
\end{figure}

In order to apply the general formula  (\ref{calX-calXbar}) we need
setting $D_1$ as the last brane on the real positive axis therefore
our setup is with $D_3$ for $x_3=-\infty<x<x_2=0$, $D_2$ for $x_2=0<x<x_1$
and $D_1$ for $x_1<x$, i.e. we have the boundary conditions
\begin{align}
&
Re(e^{-\pi \alpha_3} \partial_y X^z)=0,~~~~
Im(e^{-\pi \alpha_3} X^z)= g_3,~~~~
x_3=-\infty<x<x_2=0
\nonumber\\
&
Re(e^{-\pi \alpha_2} \partial_y X^z)=0,~~~~
Im(e^{-\pi \alpha_2} X^z)= g_2,~~~~
x_2=0<x<x_1
\nonumber\\
&
Re(e^{-\pi \alpha_1} \partial_y X^z)=0,~~~~
Im(e^{-\pi \alpha_1} X^z)= g_1,~~~~
x_1<x
\end{align}
The setup is shown in  figure \ref{fig:N3_in_out_twisted}.

\begin{figure}[hbt]
\begin{center}
\def\svgwidth{150px}
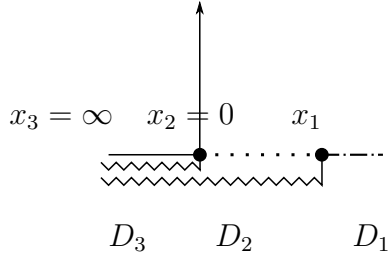
\end{center}
\vskip -0.5cm
\caption{The cuts of the classical and quantum basis.
}
\label{fig:N3_in_out_twisted_cuts_t1}
\end{figure}

A basis for the derivative of the classical solution 
on $\C-[x_N,x_1]=\C-(-\infty,x_1]$ is given by
\begin{align}
\partial\chi_{cl, n}(z)
&=
z^{-\bep_2-n-1} (z-x_1)^{-\bep_1}
\nonumber\\
\partial\bar\chi_{cl, m}(z)
&=
e^{-i 2\pi \alpha_1}
z^{-\epsilon_2-m-1} (z-x_1)^{-\epsilon_1}
,
\label{N=3-cl-basis}
\end{align}
where the power of the cut at $z=x_1$ is dictated by the boundary
conditions and the finiteness of the contribute from the area around
$u=x_1$ to the Euclidean action $S_E$.  
The same requirement from $u=0$ implies $n=m=1$.
Since $\sum_{t=1}^3 \epsilon_t=M=1$ only $\partial\chi_{cl, 0}$ has
then a finite Euclidean action.
The cuts of these basis elements are pictured in figure  
\ref{fig:N3_in_out_twisted_cuts_t1}.

Then the classical solution is given by\footnote{
the phase of $(-x_1)^{-\bep_1}$ depends on the consider $z\in H$ in
which case $(-x_1)^{-\bep_1}=(x_1)^{-\bep_1} e^{-i \pi \bep_1}$ or
$z\in H^-$ for which $(-x_1)^{-\bep_1}=(x_1)^{-\bep_1} e^{+i \pi \bep_1}$.}
\begin{align}
\chi_{cl}(z)
&= 
f_2+
c_0 \int_{w=0}^z d w w^{-\bep_2} (w-x_1)^{-\bep_1}
\nonumber\\
&=
f_2+
c_0 \frac{c_0 (-x_1)^{-\bep_1} }{\epsilon_2} z^{\epsilon_2}
{}_2F_1(\epsilon_2, \bep_1; 1+ \epsilon_2; \frac{z}{x_1})
,
\end{align}
where the constant $c_0$ is fixed by the condition
$\chi_{cl}(x_1)=f_1$.

We can now try to use the integrated eq.s (\ref{N=3-cl-basis}) 
to find a basis for the
in quantum fluctuations, i.e. for the fluctuations which behave as the
solutions (\ref{cX-usual-N=2-twisted}) as $z\rightarrow 0$.
So we can  for example write
\begin{equation}
\chi_{n}(z) \sim \int_{x_1}^z d w ~\partial\chi_{cl, n}(w)
,
\label{integrated-class-sols}
\end{equation}
where the lower integration extreme is chosen to fulfill the quantum boundary
condition $\chi_{cl}(x_1)=0$ which can  be understood as
the fact that when $g_t=0$ for $t$s all branes intersect at the origin,
i.e. $f_t=0$ for the quantum fluctuations. 

The integrals with $n\ge 1$ behave as 
$const_{n,0} + const_{n,1}~ z^{-n-\bep_2}$ for $z\rightarrow0$ .
This is not the desired behavior for $z\rightarrow 0$ because of
$const_{n,0}$. 
The required behavior needs $const_{n,0}=0$ as follows from
eq. (\ref{cX-usual-N=2-twisted}) for $z\rightarrow 0$. 
Therefore we must subtract the constants $const_{n,0}$. 
This can be achieved by making linear combinations in many different
ways, for example
between $\partial\chi_{cl, n}(w)$ with $n>1$ and $\partial\chi_{cl,  n=1}(w)$.
This approach produces a basis but gives the impression that we are missing the
$n=1$ mode.
Even if we choose a different way of substracting the constants $const_{n,0}$
we get one constraint and hence one mode ``less'' than expected.

To investigate better this point we  
start directly with a basis\footnote{There are other possible basis but this
 seems to be the most natural.} for the quantum fluctuations which
satisfies all the boundary conditions given by 
\begin{align}
\xi_{q, n}(z) 
&= 
\frac{1}{n+\bep_2} z^{-\bep_2-n}
\left(\frac{z}{x_1}-1\right)^{\epsilon_1} 
\nonumber\\ 
\bar\xi_{q, n}(z)
&= 
e^{-i 2\pi \alpha_1}
\frac{1}{n+\epsilon_2} z^{-\epsilon_2-n}
\left(\frac{z}{x_1}-1\right)^{\bep_1}
,
\label{N=3-twist-in-twist-out-cX-naive}
\end{align}
where cuts are running from $-\infty$ to either $x_2=0$ or to $x_1$
as in figure \ref{fig:N3_in_out_twisted_cuts_t1}.
They are located in this way in order to satisfy the boundary conditions. 
To these functions defined on $\C-[x_N,x_1]=\C-(-\infty,x_1]$ 
correspond the modes defined on the upper half plane $H$
\begin{align}
Y_{(c) n}(u,\bu)
=
\vect{
\frac{u^{-\bep_2-n}}{n+\bep_2}
\left(\frac{u}{x_1}-1\right)^{\epsilon_1}
}{
e^{i 2\pi \alpha_1}
\frac{\bu^{-\bep_2-n}}{n+\bep_2}
\left(\frac{\bu}{x_1}-1\right)^{\epsilon_1}
},~~~~
\bY_{(a) n}(u,\bu)
=
\vect{
\frac{\bu^{-\epsilon_2-n}}{n+\epsilon_2}
\left(\frac{\bu}{x_1}-1\right)^{\bep_1}
}{
e^{i 2\pi \alpha_1}
\frac{u^{-\epsilon_2-n}}{n+\epsilon_2}
\left(\frac{u}{x_1}-1\right)^{\bep_1}
}
.
\label{naive-N3-X-basis}
\end{align}
Notice that it is not immediate to use this basis to find the
classical solution since all these modes have infinite Euclidean action and
only a combination of them reproduce the classical solution.
In particular
\begin{align}
\partial \xi_{q, n}(z) 
=
x_1^{\bep_1} \left( 
\partial \chi_{cl, n}(z)
-
\frac{n-\epsilon_3}{n+\bep_2} \frac{1}{x_1} 
\partial \chi_{cl, n-1}(z)
\right)
\label{naive-N3-calX-basis}
\end{align}
implies that the derivative of the classical solution
$\partial \chi_{cl, 1}(z)$ is given by an infinite sum of quantum
basis elements.
\COMMENTO{{\bf wrong :( }
It also suggests how a good basis for quantum fluctuations can be
obtained from the classical basis: this suggestion will be used when
writing the expression for the zero modes in the case where the in
and out strings are untwisted in eq. (\ref{N-in-out-untwisted-zm}).
}
We now compute the non vanishing products of two quantum basis
elements and get
\begin{align}
(Y_{(c) n}, \bY_{(a) m} )
& 
= \left[ (\bY_{(a) m}, Y_{(c) n} ) \right]^*
=
\frac{
-4\pi i
}{
(n+\bep_2)
}
\left[
-\delta_{n+m,-1}
+ 
\frac{m+\epsilon_3}{m+\epsilon_2}
\frac{1}{x_1} \delta_{n+m,0}
\right]
.
\end{align}
This means that these modes are not orthogonal and we need to find 
combinations which are orthogonal. 
For any $B\in \Z$ we can then search orthogonal solutions in the form
\begin{align}
\chi_{q, n}(z) 
&= 
\left\{
\begin{array}{c r}
\left( \sum_{k=-B}^n
c_{n\, k} \frac{1}{n+\bep_2} z^{-\bep_2-n} \right)
\left(\frac{z}{x_1}-1\right)^{\epsilon_1} 
& n  \ge -B
\\
\frac{1}{n+\bep_2} z^{-\bep_2-n}
\left(\frac{z}{x_1}-1\right)^{\epsilon_1} 
 & n \le - B -1
\end{array}
\right.
\nonumber\\ 
\bar\chi_{q, n}(z)
&= 
\left\{
\begin{array}{c r}
e^{-i 2\pi \alpha_1}
\left( \sum_{k=-\infty}^n
\bc_{n\, k} \frac{1}{n+\epsilon_2} z^{-\epsilon_2-n} \right)
\left(\frac{z}{x_1}-1\right)^{\bep_1}
& n \ge B
\\
e^{-i 2\pi \alpha_1}
\frac{1}{n+\epsilon_2} z^{-\epsilon_2-n}
\left(\frac{z}{x_1}-1\right)^{\bep_1}
& n \le B-1
\end{array}
\right.
.
\label{N=3-twist-in-twist-out-cX-ortho}
\end{align}
The reason of such an ansatz is that given the integer $B$,
$\chi_{q,n}=\xi_{q,n}$ with $n\le -B-1$ and $\bar \chi_{q,m}=\bar
\xi_{q,m}$  with $m\ge B-1$
are automatically orthogonal and are the biggest set of orthogonal
elements  among the original $\xi$ and $\bar\xi$.
The explicit solution is then
\begin{align}
c_{n\, k} &= \left(\frac{1}{x_1}\right)^{n-k} 
\frac{ (-n-\epsilon_3)_{n-k} }{ (-n-\bep_2)_{n-k} }
c_{n\, n}
\nonumber\\
\bc_{n\, k} &= \left(\frac{1}{x_1}\right)^{n-k} 
\frac{ (-n+\epsilon_3)_{n-k} }{ (-n-\epsilon_2)_{n-k} }
\bc_{n\, n}
.
\end{align}
In particular we recognize  that for $ n \ge B$
\begin{align}
\bar\chi_{q, n}(z)
&=
e^{-i 2\pi \alpha_1}
\frac{1}{n+\epsilon_2} z^{-\epsilon_2-n}
\left(\frac{z}{x_1}-1\right)^{\bep_1}
~{}_2F_1(1, -n+\epsilon_3; -n+\bep_2; \frac{z}{x_1})
\end{align}
when we set $\bc_{n\, n}=1$.
This is one of the solution associated with the Papperitz- Riemann
symbol of the hypergeometric
\begin{align}
z^{-n-\epsilon_2}
\left(\frac{z}{x_1}-1\right)^{\bep_1}
P\left\{
\begin{array} {c c c c}
0 & 1 & \infty & \\
0 & 0 & 1 & \frac{z}{x_1} \\
n+\epsilon_2 & -\bep_1 & -n+\epsilon_3 & 
\end{array}
\right\}
=
P\left\{
\begin{array} {c c c c}
0 & 1 & \infty & \\
0 & 0 & 0 & \frac{z}{x_1} \\
-n-\epsilon_2 & \bep_1 & n+1-\bep_3 & 
\end{array}
\right\}
.
\label{Pap-Rie}
\end{align}
The last form of the $P$ symbol clearlu shows that the indeces around
the singular points are the desired ones.
The $P$ symbols also shows that the other solution is simply 
the constant $1$.
This is important when performing the analytic continuation
of the hypergeometric to the region $ | z /x_1 |> 1$.
In fact in the region  $ | z /x_1 |> 1$ we find that the two
independent solutions mix and the analytically continued solution does not
satisfy anymore the required boundary conditions.
This means that we need again to consider combinations of the
solutions found above in the $ | z /x_1 |<1$  so that their
continuation satisfies the proper boundary conditions.
Therefore we find again one mode less than expeceted.

Another way of seeing the problem is to notice that the hypergeometric
equation associated with the symbol (\ref{Pap-Rie}) has 
$a=0, b=n+1-\bep_3$ and $c=n+1+\epsilon_2$ so that the hypergeometric
equation reads
\begin{equation}
w (1-w)\frac{d^2 \bar\chi_{q, n}(w)}{ d w^2} +
\left[ n+1+\epsilon_2 -( n-\bep_3 +2) w\right]
 \frac{d \bar\chi_{q, n}(w)}{ d w}=0
\end{equation}
with $w=z/x_1$.
This can immediately integrated and gives
\begin{align}
\bar\chi_{q, n}(z)
&=
e^{-i 2\pi \alpha_1}
\int^z_{x_1} d w~
w^{-\epsilon_2-n-1}
\left(\frac{w}{x_1}-1\right)^{-\epsilon_1},
~~~
n\ge B
\label{non-simple-chis}
\end{align}
exactly as for the classical solutions in eq. (\ref{integrated-class-sols}).
It has therefore the same problems with the boundary conditions as before.
This problem can be generalized to the generic $N$ and it is that one
asymptotic mode (either for $u\rightarrow0$ or equivalently for
$u\rightarrow \infty$) is missing for any twist field we insert at
$x_t\ne 0, \infty$.
This is caused by the constraints $X_q(x_t,x_t)=0$ we have to impose
on the quantum fields.
Physically we can partially understand this as a breaking of some
symmetries, for example if the in string is untwisted the momentum
conservation of the whole is broken because the other branes at angles.
Another naive way of understanding is that the classical solution freezes
these modes.
It would however be nice to have a better understanding of this problem and of
how to choose the basis.

From the previous discussion we learn however a rule how to write a system
of orthogonal modes for a generic $N$. The rule is simple 
up to the further linear
combinations/subtractions in order to satisfy the boundary
conditions in the whole upper plane when performing the analytic continuation.
We start from a maximally orthogonal subset of the would be quantum
modes $\chi_q$ and $\bar\chi_q$. 
This subset corresponds to the modes 
$\chi_{q,n}=\xi_{q,n}$ with $n\le -B-1$ and $\bar \chi_{q,m}=\bar
\xi_{q,m}$  with $m\ge B-1$ in the case of this section.
Since the product is roughly 
$\oint dz \chi_q(z) \frac{d \bar \chi_q}{d z}$, we can define the remaining 
orthogonal modes by choosing their derivatives 
$\frac{d \bar \chi_q}{d  z}$ so that the products 
$\chi_q(z) \frac{d \bar \chi_q}{d z}$ have only simple poles. 
This practically means that 
$\frac{d \bar \chi_q}{d  z} \sim \frac{d \bar \chi_{cl}}{d  z}$.
In the case of this section in the integrated form this corresponds to
the modes given in eq.(\ref{non-simple-chis}).
\subsection{Normal modes and improved overlap approach}
\label{sect:normal_modes_overlap}
The issue is how to cope with the problem in finding  
a basis of orthogonal modes with the proper boundary conditions
and havig all the possible asymptotic modes.
From the discussion in the previous section this is probably impossible.
The simplest way to proceed further is to consider the standard
approach in quantum mechanics.
The standard procedure in quantum
mechanics in presence of discontinuities is using the overlap
of the wave function and that the basis of the incoming wave functions is
complete and orthogonal.
This is essentially the same approach used by Cremmer, Gervais, Kato,
Ogawa and Mandelstam in the early seventies in computing the
three-string vertex (\cite{Goto:1974uq}, \cite{overlap}) and the
extended to the case of the $D1$ string in \cite{Pesando:2013qda}.

There is however a slight subtlety to be stressed and understood 
before we can apply the overlap method to our case.
In previous papers there were three strings and the conditions imposed
were the overlap of the string coordinates $X$  and their momenta
$P$.
In our case we have only one string with discontinuous boundary conditions
but we still want to impose the continuity of the coordinate.
In the case $N=3$ of section \ref{sect:N=3-in-out-twisted} we could
think of simply using the in and out expansions and require at the
transition point
$
X^{(in)}(x_1,x_1)=X_{cl}^{(in)}+X_{q}^{(in)}(x_1,x_1)
=
X^{(out)}(x_1,x_1)=X_{cl}^{(out)}+X_{q}^{(out)}(x_1,x_1)
$
with $X_{cl}$ the constant given in eq. (\ref{N=2-X-cl}) 
and $X_q(u,\bu)$ the quantum field given in eq. (\ref{loc-exp-X}) for
in and out strings.
Nevertheless in the case where multiple transitions 
are encountered 
as we discuss in the next section
it is not clear which is the classical part of the intermediate strings.
We are therefore led to proceed as follows.
First we split $X(u,\bu)$ into the classical solution $X_{cl}(u,\bu)$
and the quantum part $X_q(u,\bu)$.
The classical solution has a global nature and must be computed for
the whole evolution before proceeding to the second step.
Second we require that only the quantum fluctuations overlap.
This approach is a refined version of the overlap approach and is 
consistent with the naive idea of overlap of the
whole string coordinate.

We formulate therefore the overlap approach as\footnote{
As done in the whole paper we suppose  that all $x_t>0$.}
\begin{equation}
X_{q}^{(t+1)}( u, \bu)\Big|_{|u|=x_t-0^+}
=
X_{q}^{(t)}( u, \bu)\Big|_{|u|=x_t+0^+}
\label{gen-overlap}
\end{equation}
where $X_{q}^{(t)}( u, \bu)$ is the quantum fluctuation of the
string comprised in the half annulus with $x_{t}<|u|< x_{t-1}$ and
having the appropriate boundary conditions.

\section{Green function, in and out vacuum}
\label{Green_function_in_out_vacua}
\begin{figure}[hbt]
\begin{center}
\def\svgwidth{300px}
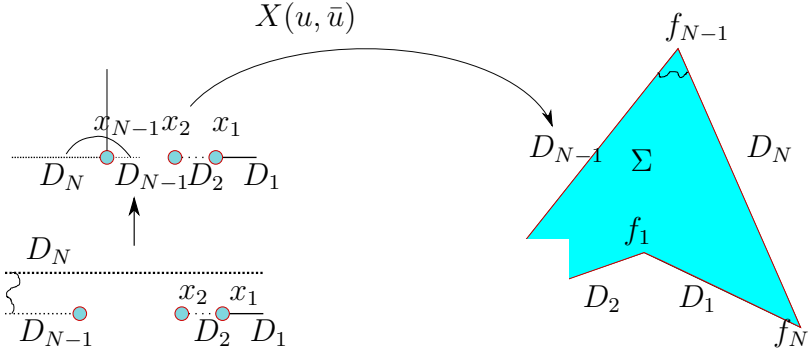
\end{center}
\vskip -0.5cm
\caption{Map from the stripe to the upper half plane to the target polygon
  $\Sigma$ with twisted in and out strings.}
\label{fig:H2polygon_twisted_in_out}
\end{figure}

From now on we adopt the strategy outlined in the previous section and
focus on the configuration pictured in figure
\ref{fig:H2polygon_twisted_in_out}  where both in and out states are
twisted. 
This means that $x_{N-1}=0$ and $x_{N}=\infty$. 

In principle we should be able to derive the Green function from the
canonical quantization. 
The derivative of the Green function is given in operatorial formalism
by
\begin{align}
\partial_u \partial_v G^{I J}(u, \bu; v, \bv; \PaMoSp)
=
\frac{
\langle 0_{out}| 
\partial_u X^I_q(u,\bu) \partial_v X^J_q(v,\bv)
|0_{in}\rangle
}{
\langle 0_{out}|0_{in}\rangle 
}
\label{Green-oper}
\end{align}
but since we have to yet completely understood
the global modes we assume we compute the Green function in the usual
way i.e. using its singularities and boundary conditions.
Notice that in the configuration we consider there is implicit  the limit
$x_N\rightarrow \infty$, however  this limit does not require any
special treatment or factor since the CFT definition of the derivative
of the Green function gives
\begin{align}
\partial_u \partial_v G^{I J}(u, \bu; v, \bv; \PaMoSp)|_{x_N=\infty}
=
\lim_{x_N\rightarrow \infty}
\frac{
x_N^{ \epsilon_N \bep_N}
\langle 
\partial_u X^I_q(u,\bu) \partial_v X^J_q(v,\bv)
\prod_{t=1}^N \sigma_{\epsilon_t}(x_t)
\rangle
}{
x_N^{ \epsilon_N \bep_N}
\langle \prod_{t=1}^N \sigma_{\epsilon_t}(x_t) \rangle 
}
.
\label{Green-abs}
\end{align}
Comparison between the two previous expression (\ref{Green-oper}) and
(\ref{Green-abs}) suggest that we can identify
\begin{equation}
\langle 0_{out}|0_{in}\rangle 
=
\lim_{x_N\rightarrow \infty}
x_N^{ \epsilon_N \bep_N}
\langle \prod_{t=1}^N \sigma_{\epsilon_t}(x_t) \rangle
\end{equation}
for the configuration we consider.
In the following we will assume such a natural identification.

Given the Green function we can consider the region $|u|<x_{N-2}$. 
Using the overlap approach we can identify
$\partial_u X^I_q(u,\bu)$ with the twisted string 
$\partial_u X^I_{T(D_{N-1}\, D_N), q}(u,\bu)$. 
This twisted string has the boundary conditions which follow from
being attached to the $D_{N-1}$ brane on $x>0$ and 
to the $D_{N}$ brane on $x<0$.
The derivative of the Green function can then be written as
\begin{align}
\partial_u \partial_v & G^{I J}(u, \bu; v, \bv; \PaMoSp)
\nonumber\\
&=
\frac{
\langle 0_{out}| 
\partial_u X^I_{T(D_{N-1}\, D_N), q}(u,\bu) 
\partial_v X^J_{T(D_{N-1}\, D_N), q}(v,\bv)
|0_{in}\rangle
}{
\langle 0_{out}|0_{in}\rangle 
}
,~~~~
|u|,|v|<x_{N-2}
.
\end{align}
In particular we can identify the in vacuum with the vacuum of  the
twisted Hilbert space $\cH_{(D_{N-1} D_N)}$
\begin{equation}
|0_{in}\rangle = | T_{T(D_{N-1}\, D_N)} \rangle
.
\end{equation}
After performing the normal ordering with respect to this vacuum
the previous equation can then be written as
\begin{align}
\partial_u \partial_v &\Delta^{I J}_{(N,M)(t=N-1)}(u, \bu; v, \bv; \PaMoSp)
\nonumber\\
&=
\frac{
\langle 0_{out}| 
:\partial_u X^I_{T(D_{N-1}\, D_N), q}(u,\bu) 
\partial_v X^J_{T(D_{N-1}\, D_N), q}(v,\bv):
|0_{in}\rangle
}{
\langle 0_{out}|0_{in}\rangle
}
,
\label{der-der-Delta-t=N-1}
\end{align}
where we have defined the Green function regularized at the twisted
interaction point $x_t$ with $t=N-1$ (in the case at hand $x_{N-1}=0$)
in the sector $M=\sum_{t=1}^N \epsilon_t$ as
\begin{align}
\Delta^{I J}_{(N,M)(t=N-1)}(u, \bu; v, \bv; \PaMoSp)
=&
G^{I J}_{(N,M)}(u, \bu; v, \bv; \PaMoSp)
\nonumber\\
&-
G^{I J}_{N=2, T(D_{N-1}\, D_N)}(u, \bu; v, \bv; \PaMoSpNDuenn{N-1}{N})
,
\label{DeltaNM-t=N-1}
\end{align}
where $G^{I J}_{N=2, T(D_{N-1}\, D_N)}$ is the twisted Green function
given in eq. (\ref{Green-N=2}) for the boundary conditions associated
to the Hilbert space $\cH_{(D_{N-1} D_N)}$, i.e. with 
the twist $\sigma_{\epsilon_{N-1},f_{N-1}}$  at $x=0$ and 
the anti-twist  $\sigma_{\bar \epsilon_{N-1},f_{N-1}}$ in $x=\infty$.

From equation (\ref{der-der-Delta-t=N-1}) we can then determine the
coefficients which enter the operatorial expression for the out vacuum
\begin{align}
\langle 0_{out}|
=
\cN
\langle 0_{in}|
\exp\Big\{
&
\oh \sum_{n,m=0}^\infty \Big[
B_{n \bz, m \bz} \alpha_{n+\epsilon_{N-1}} \alpha_{m+\epsilon_{N-1}}
+
B_{n z, m z} \bar\alpha_{n+\bep_{N-1}} \bar\alpha_{m+\bep_{N-1}}
\nonumber\\
&
+
2 B_{n \bz, m z} \alpha_{n+\epsilon_{N-1}} \bar\alpha_{m+\bep_{N-1}}
\Big]
\Big\}
\end{align}
with $B_{n \bz, m z}= B_{m z, n \bz}$.
A simple computation gives $\cN=\langle 0_{out}|0_{in}\rangle$.
If we take $I J = z z$ in eq. (\ref{der-der-Delta-t=N-1}) we get the 
$\partial_u\partial_v$ derivative of
\begin{align}
\sum_{n,m=0}^\infty 
&
\Big[
B_{n \bz, m \bz} u^{n+\bep_{N-1}} v^{m+\bep_{N-1}}
+
B_{n z, m z} \bu^{n+\bep_{N-1}} \bv^{m+\bep_{N-1}}
\nonumber\\
&
+
B_{n \bz, m z} ( u^{n+\epsilon_{N-1}} \bv^{m+\bep_{N-1}} + v^{n+\epsilon_{N-1}} \bu^{m+\bep_{N-1}})
\Big]
=
\nonumber\\
&=
\left( -i \frac{\sqrt{2\alpha'}}{2} e^{i \pi \alpha_{N-1}}
\right)^{-2}
\Delta^{z z}_{(N,M)(t=N-1)}(u, \bu; v, \bv; \PaMoSp)
.
\label{Deltazz}
\end{align}
If we consider the analogous expressions of
eq. (\ref{der-der-Delta-t=N-1}) with the derivatives $\partial_u\partial_\bv$,
$\partial_\bu\partial_v$ and
$\partial_\bu\partial_\bv$ we find the corresponding derivatives of
the previous equation (\ref{Deltazz}).
The derivative expressions can be integrated to obtain eq. (\ref{Deltazz})
where the integration constant is fixed to be zero because of the cut structure.
Similarly if we take $I J = z \bz$ we get the derivatives of
\begin{align}
\sum_{n,m=0}^\infty 
&
\Big[
B_{n \bz, m \bz} u^{n+\bep_{N-1}} \bv^{m+\bep_{N-1}}
+
B_{n z, m z} \bu^{n+\bep_{N-1}} v^{m+\bep_{N-1}}
\nonumber\\
&
+
B_{n \bz, m z} ( u^{n+\epsilon_{N-1}} v^{m+\bep_{N-1}} + \bv^{n+\epsilon_{N-1}} \bu^{m+\bep_{N-1}})
\Big]
=
\nonumber\\
&=
\left( -i \frac{\sqrt{2\alpha'}}{2}
\right)^{-2}
\Delta^{z \bz}_{(N,M)(t=N-1)}(u, \bu; v, \bv; \PaMoSp)
,
\end{align}
where the integration constant is again fixed to be zero 
because of the cut structure.
There is also the third possibility given by  $I J = \bz \bz$.
Nevertheless all three expressions for the $B$ coefficients 
are equivalent because of the $\Delta^{I J}$ boundary conditions.

The previous equation can also be rewritten in a more compact form as
\begin{align}
\langle 0_{out}|
=
\cN
\langle 0_{in}|
\exp\Big\{
\oint_{z=0} \frac{d z}{ 2\pi i} \oint_{w=0} \frac{d w}{ 2\pi i} 
&
\partial_z \cX^{I (+)}_{T(D_{N-1}\, D_N), q}(z) 
\partial_w \cX^{J (+)}_{T(D_{N-1}\, D_N), q}(w)
\nonumber\\
&
\Delta^{I J}_{(N,M) (t=N-1) L L}(z; w; \PaMoSp)
\Big\}
,
\end{align}
where $\Delta^{I J}_{(N,M) (t=N-1) L L}(z; w)$ is the holomorphic part
in both $u$ and $v$ analytically continued in the complex plane minus cuts.

Finally let us comment on why we have not explicitly indicated the
$x_N\rightarrow \infty$ limit in all the previous expressions.
The reason is that we could repeat all the previous arguments for a
configuration with a finite $x_N$.
In particular also for an outgoing untwisted vacuum.
There would be two minor differences.
The one difference with the previous results is not taking 
the $x_N\rightarrow \infty$ limit.
The second one is the relation between the
correlator of the plain twist fields
 and the product of the in and out vacua which would read
\begin{equation}
\langle 0_{out}|0_{in}\rangle 
=
\langle \prod_{t=1}^N \sigma_{\epsilon_t}(x_t) \rangle
~~~~\mbox{with finite }~ x_N
.
\end{equation}

\section{The generating function of amplitudes}
\label{reggeons}
In this section we would like to perform the actual computations of the
generating functions for amplitudes involving  plain and excited
twisted states. This is done in steps.  First considering the
amplitudes with plain unexcited twisted fields and arbitrary untwisted
states. Then considering amplitudes with excited twisted states
without untwisted ones and finally, assembling all.

\subsection{Amplitudes with untwisted vertices and plain twist fields}
\label{reggeon_untwisted}
Given the previous knowledge of the out vacuum as a sliver constructed
on the in vacuum
we can now easily compute boundary amplitudes with vertices in the $|u|<x_{N-1}$
region.
In particular we want to compute the generating function of the amplitudes
\begin{equation}
\langle \prod_{i=1}^L V_{\xi_i}(\hx_i) 
\prod_{t=1}^N \sigma_{\epsilon_t, f_t}(x_t) \rangle
\label{sample-L-untw-correlator}
\end{equation}
with $|\hx_i|<x_{N-1}$. 
$\xi_i$ is a generic untwisted state living on $D_{N-1}$
or $D_N$ and $V_{\xi_i}(\hx_i)$ its emission vertex.
To  this end we need mapping the abstract vertex operator
$V_{\xi_i}(\hx_i)$  into  its operatorial realization in the twisted
Hilbert space $\cH_{T(D_{N-1} D_N)}$.
This mapping is realized using the SDS vertex $\cS(x; \{c^I\})$.
The SDS vertex for the twisted
Hilbert space $\cH_{T(D_{N-1} D_N)}$ is explicitly given by
\COMMENTO{
\begin{align}
\cS_{T(D_{N-1} D_N)}(c^I)
=
&:
\exp\left\{
\sum_{n=0}^\infty  
\left[ \bar c_n \partial^n_\xl X_{(D_{N-1} D_N)}(\xl+i0^+,\xl-i0^+)
+ c_n \partial^n_\xl \bar X_{(D_{N-1} D_N)}(\xl+i0^+,\xl-i0^+) \right]
\right\}
:
\nonumber\\
&
\exp\left\{
\oh
\sum_{n,m=0}^\infty  
\bar c_n~c_m ~\partial^n_\xii1 ~\partial^m_\xii2
\Delta\bou^{z \bz}(\xii1; \xii2; \PaMoSpNDuenn{N-1}{N})
|_{\xii1=\xii2=\xl} 
\right\}
\nonumber\\
&
\exp\left\{
\oh
\sum_{n,m=0}^\infty  
c_n~c_m ~\partial^n_\xii1 ~\partial^m_\xii2
\Delta\bou^{\bz \bz}(\xii1; \xii2; \PaMoSpNDuenn{N-1}{N}) 
|_{\xii1=\xii2=\xl} 
\right\}
\nonumber\\
&
\exp\left\{
\sum_{n,m=0}^\infty  
\bar c_n~\bar c_m ~\partial^n_\xii1 ~\partial^m_\xii2 
\Delta\bou^{z z}(\xii1; \xii2; \PaMoSpNDuenn{N-1}{N})
|_{\xii1=\xii2=\xl} 
\right\}
\label{S_bou}
\end{align}
} 
\begin{align}
\cS_{T(D_{N-1} D_N)} & (x; \{c^I\})
=
:
\exp\left\{
\sum_{n=0}^\infty  
c_{n I}~\partial^n_x X^I_{T(D_{N-1} D_N)}(x+i0^+, x-i0^+)
\right\}
:
\nonumber\\
&
\exp\left\{
\oh
\sum_{n,m=0}^\infty  
c_{n I}~c_{m J} ~\partial^n_\xii1 ~\partial^m_\xii2
\Delta_{T(D_{N-1} D_N), bou(x)}^{I J}(\xii1; \xii2; \PaMoSpNDuenn{N-1}{N})
|_{\xii1=\xii2=x} 
\right\}
,
\label{S_bou}
\end{align}
where $:\dots:$ is the normal ordering in the 
twisted Hilbert space. 
We defined the boundary regularized Green function for the twisted
Hilbert space  at the point $x$
\begin{align}
\Delta_{T(D_{N-1} D_N), bou(x)}^{I J}
&
(\xii1; \xii2; \PaMoSpNDuenn{N-1}{N})
\nonumber\\
=&
G^{I J}_{N=2, T(D_{N-1}\, D_N)}(\xii1+i0^+,\xii1-i0^+; \xii2+i0^+,\xii2-i0^+; 
\PaMoSpNDuenn{N-1}{N} ) 
\nonumber\\
&
- G^{I J}_{U(t_x)}(\xii1+i0^+,\xii1-i0^+; \xii2+i0^+,\xii2-i0^+; \alpha_{N-1})
,
\label{DeltaTN-1N}
\end{align}
where $t_x$  is the index of the brane on which the vertex with
coordinate $x$ is. 
For $x_t < x< x_{t-1}$ we have $t_x=t$. 
In the case at hand we have actually only two possibilities, either
$0<x<x_{N-1}$ then $t_x=N-1$ or $-x_{N-1}< x < 0 $ then $t_x=N$.

In this expression 
$G^{I J}_{N=2, T(D_{N-1}\, D_N)}(u,\bu; v,\bv; \PaMoSpNDuenn{N-1}{N})$ 
the same Green function used in eq. (\ref{DeltaNM-t=N-1})
and given in eq. (\ref{Green-N=2}).
$ G^{I J}_{U(t_x)}(u,\bu; v,\bv)$ is the Green function of the
untwisted string with both ends on $D_{t_x}$ and 
given in eq. (\ref{untwisted-green-function}).

Then we can realize the mapping as
\begin{align}
\left[
e^{i \bk X + i k \bar X}
\prod_{n=1}^\infty 
\left(\partial^n_x X \right)^{N_n}
\left(\partial^n_x \bar X \right)^{\bar N_n}
\right](x+i0^+, x-i 0^+)
\leftrightarrow
\left.
\prod_{n=1}^\infty 
\frac{\partial^{N_n}}{\partial \bar c_n^{N_n}}
\frac{\partial^{\bar N_n}}{\partial c_n^{\bar N_n}}
\cS(x; \{c,\bc\}) \right|_{c_{0 I}=i\,k_I,c_{n\ge1}=0}
\end{align}
when we identify $c_z=c^\bz=c$ and $c_\bz=c^z=\bar c$.
This is the correct map because operatorial realizations have the same
OPEs as the abstract operators  despite the use of the twisted Green
function associated with the  twisted fields $X^I_{T(D_{N-1} D_N)}$.
It is worth stressing for what follows that the fields
 $X^I_{T(D_{N-1} D_N)}$ are the
full fields and not only the quantum fluctuations.

In order to compute the generating function of all the correlators
like (\ref{sample-L-untw-correlator})
it is then enough to insert a SDS for any  untwisted operator into the
radial ordered  expression
\begin{align}
V_{0+L}( \{\hx_i;\{c_{(i)}\}\})=
\langle 0_{out}| 
R\left[ \prod_{i=1}^L \cS_{(D_{N-1} D_N)}(\hx_i; \{c_{(i)}\})  \right]
| 0_{in}\rangle
.
\label{L-Untwisted-Reggeon-restricted}
\end{align}
To compute explicitly the previous expression we have to split
the full fields  $X^I_{T(D_{N-1} D_N)}$ into classical and quantum
parts and then normal order the quantum parts.
After these operations we get
\begin{align}
V_{0+L}( \{\hx_i; 
&\{c_{(i)}\}\})
=
\nonumber\\
&
\prod_{i=1}^L
\Bigg\{
\exp\left[
\sum_{n=0}^\infty  
c_{(i) n I}~\partial^n_x X^I_{cl}(\hx_i+i0^+, \hx_i-i0^+; \PaMoSpFull)
\right]
\nonumber\\
&
\exp\left[
\oh
\sum_{n,m=0}^\infty  
c_{(i) n I}~c_{(i) m J} ~\partial^n_\xii1 ~\partial^m_\xii2
\Delta_{T(D_{N-1} D_N), bou(i)}^{I J}(\xii1; \xii2; \PaMoSpNDuenn{N-1}{N})
|_{\xii1=\xii2=\hx_i} 
\right]
\Bigg\}
\nonumber\\
&
\prod_{1\le i<j \le L}
\exp\left[
\sum_{n,m=0}^\infty  
c_{(i) n I}~c_{(j) m J} ~\partial^n_\xii1 ~\partial^m_\xii2
G_{N=2, T(D_{N-1}\, D_N)}^{I J}(\xii1, \xii1; \xii2, \xii2; \PaMoSpNDuenn{N-1}{N})
|_{\xii1=\hx_i; \xii2=\hx_j} 
\right]
\nonumber\\
&
\langle 0_{out}|
\exp\left[
\sum_{i=1}^L
\sum_{n=0}^\infty  
c_{(i) n I}~\partial^n_x X^{I (-)}_{T(D_{N-1}\, D_N),q}(\hx_i+i0^+, \hx_i-i0^+)
\right]
|0_{in}\rangle
,
\end{align}
where $bou(i)$ is a short hand for $bou({\hx_i})$.
The term in the last line can be evaluated using the analogous
expression of $ \langle0| e^{\beta a^2} e^{\alpha a^\dagger} |0\rangle
= e^{ \beta \alpha^2}$ for an infinite number of
oscillators and gives
\begin{align}
\langle 0_{out}|
&
\exp\left[
\sum_{i=1}^L
\sum_{n=0}^\infty  
c_{(i) n I}~\partial^n_x X^{I (-)}_{T(D_{N-1}\, D_N), q}(\hx_i+i0^+, \hx_i-i0^+)
\right]
|0_{in}\rangle
\nonumber\\
=&
\lim_{x_N\rightarrow \infty}
x_N^{ \epsilon_N \bep_N}
\langle \prod_{t=1}^N \sigma_{\epsilon_t, f_t}(x_t) \rangle
\nonumber\\
&\times
\prod_{1\le i,j \le L}
\exp\left[
\oh
\sum_{n,m=0}^\infty  
c_{(i) n I}~c_{(j) m J} ~\partial^n_\xii1 ~\partial^m_\xii2
\Delta_{(N,M) (t=N-1)}^{I J}(\xii1, \xii1; \xii2, \xii2; \PaMoSp)
|_{\xii1=\hx_i; \xii2=\hx_j} 
\right]
,
\end{align}
where $\Delta_{(N,M) (t=N-1)}^{I J}$ is the expression in
eq. (\ref{DeltaNM-t=N-1}).
When we assemble all contributions we find finally
\begin{align}
V_{0+L}( \{\hx_i; 
&\{c_{(i)}\}\})
=
\lim_{x_N\rightarrow \infty}
x_N^{ \epsilon_N \bep_N}
\langle \prod_{t=1}^N \sigma_{\epsilon_t, f_t}(x_t) \rangle
%
%
\nonumber\\
%
&
\prod_{i=1}^L
\Bigg\{
\exp\left[
\sum_{n=0}^\infty  
c_{(i) n I}~\partial^n_x X^I_{cl}(\hx_i+i0^+, \hx_i-i0^+; \PaMoSpFull)
\right]
\nonumber\\
&
\exp\left[
\oh
\sum_{n,m=0}^\infty  
c_{(i) n I}~c_{(i) m J} ~\partial^n_\xii1 ~\partial^m_\xii2
\Delta_{(N,M) bou(i)}^{I J}(\xii1; \xii2; \PaMoSp)
|_{\xii1=\xii2=\hx_i} 
\right]
\Bigg\}
\nonumber\\
%
&
\prod_{1\le i<j \le L}
\exp\left[
\sum_{n,m=0}^\infty  
c_{(i) n I}~c_{(j) m J} ~\partial^n_\xii1 ~\partial^m_\xii2
G^{I J}_{(N,M)}(\xii1, \xii1; \xii2, \xii2; \PaMoSp)
|_{\xii1=\hx_i; \xii2=\hx_j} 
\right]
,
\label{V_untwisted}
\end{align}
where we have introduced the boundary Green function in sector $M$ 
regularized at the point $\hx_i$ 
\begin{align}
\Delta_{(N,M) bou(i)}^{I J}(\xii1; \xii2; \PaMoSp)
=
&
G^{I J}_{(N,M)}(\xii1+i0^+,\xii1-i0^+; \xii2+i0^+,\xii2-i0^+; 
\PaMoSp ) 
\nonumber\\
&
- G^{I J}_{U(t_i)}(\xii1+i0^+,\xii1-i0^+; \xii2+i0^+,\xii2-i0^+;
\alpha_{t_i})
\label{Delta-N-1}
\end{align}
with $t_i=N-1$ for $\hx_i>0$ and  $t_i=N$ for $\hx_i<0$.
This is the result of the computation 
$\Delta_{(N,M) (t=N-1)}^{I J}- \Delta_{T(D_{N-1} D_N), bou(\hx_i)}^{I J}$.
The expressions of the two terms are given in eq.s (\ref{DeltaNM-t=N-1}) and
(\ref{DeltaTN-1N}). 
We have now found the generating function for the emission of
untwisted states from either $D_{N-1}$ of $D_N$ and with
$|\hx_i|<x_{N-1}$. 
Eq. (\ref{V_untwisted}) is the same expression found
in \cite{Pesando:2014owa} (eq. 80 in section 4.1) 
when we drop the $x_N\rightarrow \infty$ limit, we let the $\hx_i$ be
generic and not constrained by $|\hx_i|< x_{N-2}$ and
we substitute $\Delta_{(N,M) bou(i)}^{I
  J}(\xii1; \xii2; \PaMoSp) \rightarrow 
\Delta_{(N,M) bou(i)}^{I J}(\xii1; \xii2; \PaMoSp)$.
The expression in  \cite{Pesando:2014owa} is however 
valid without the constraint on $\hx$.
The reason for this substantial equality is quite simple.
Suppose we want compute the generating function for the emission of
untwisted states from any of the brane of the configuration we
consider. 
We need therefore computing the analogous expression of
eq. (\ref{L-Untwisted-Reggeon-restricted}) where the SDS vertices are 
the proper ones for the untwisted states.
For example, for $x_t< \hx_i< x_{t-1}$ the emission of an untwisted state 
is from the brane $D_t$.
The corresponding operatorial vertex is realized in the twisted
Hilbert space $\cH_{(D_{t} D_N)}$ using the full fields $X^I_{T(D_{t} D_N)}$.
The map abstract from operator to its operatorial realization 
is then performed by a SDS 
vertex analogous to (\ref{S_bou}) but defined in the twisted
Hilbert space $\cH_{(D_{t} D_N)}$ using the fields $X^I_{T(D_{t}  D_N)}$.

To compute the analogous expression of
eq. (\ref{L-Untwisted-Reggeon-restricted}) but with no constraints on
the location of the vertices we proceed as follows.
We split all the full fields  $X^I_{T(D_{t} D_N)}$ 
into classical and quantum parts.
Then we can use the continuity equation
(\ref{gen-overlap}) 
for the quantum part to perform an analytic continuation 
of the quantum part of any SDS vertex from an arbitrary region to
the region  $|u|<x_{N-2}$.
This seems at first sight quite difficult because of the normal
ordering which differs among the different twisted Hilbert spaces.
Fortunately it is not so.
The key observation is that the normal ordered SDS vertex in
eq. (\ref{S_bou}) is obtained from a non normal ordered vertex 
as 
\begin{align}
\cS_{T(D_{t} D_N)}(x; \{c^I\})
=
&
\lim_{\eta \rightarrow 0^+}
e^{
\sum_{n=0}^\infty  
c_{n I}~\partial^n_x 
\left[ 
X^{I(+)}_{T(D_{t} D_N)}(x+i0^+, x-i0^+)
+
X^{I(-)}_{T(D_{t} D_N)}(x e^{-\eta} +i0^+, x e^{-\eta} -i0^+)
\right]
}
\nonumber\\
&
\exp\left\{
-\oh
\sum_{n,m=0}^\infty  
c_{n I}~c_{m J} ~\partial^n_\xii1 ~\partial^m_\xii2
G_{U(t_x)}^{I J}(\xii1, \xii1; \xii2, \xii2; \alpha_{t_x} )
|_{\xii1= \xii2 e^{\eta} =x} 
\right\}
\label{S_bou_non_norm_ord}
\end{align}
with $t_x=t$ when $x>0$.
We can then easily perform the analytic continuation and get
\begin{align}
\cS_{T(D_{t} D_N)}(x; \{c^I\})
&=
\cS_{T(D_{N-1} D_N)}(x; \{c^I\})
\nonumber\\
\lim_{\eta \rightarrow 0^+}
&
e^{
-\oh
\sum_{n,m=0}^\infty  
c_{n I}~c_{m J} ~\partial^n_\xii1 ~\partial^m_\xii2
\left[
G_{U(t_x)}^{I J}(\xii1, \xii1; \xii2, \xii2; \alpha_{t_x} )
-
G_{U(N-1)}^{I J}(\xii1, \xii1; \xii2, \xii2; \alpha_{N-1} )
\right]
|_{\xii1= \xii2 e^\eta=x} 
}
,
\end{align}
where it is necessary to take the ${\eta \rightarrow 0^+}$ limit
in the last line only after we have continued back the result to the
original position.
Only then the difference $G_{U(t)}^{I J} - G_{U(N-1)}^{I J}$ which
would otherwise be divergent in the limit  ${\eta \rightarrow 0^+}$ combines
with $\Delta_{(N,M) bou(i)}^{I J}$ given in eq. (\ref{Delta-N-1}) and
valid for $|\hx|< x_{N-2}$ to
give the  finite boundary Green function regularized at
the original point $x_t<\hx< x_{t-1}$, i.e.
\begin{align}
\Delta_{(N,M) bou(i)}^{I J}(\xii1; \xii2; \PaMoSp)
=
&
G^{I J}_{(N,M)}(\xii1+i0^+,\xii1-i0^+; \xii2+i0^+,\xii2-i0^+; 
\PaMoSpNDuen{N-1} ) 
\nonumber\\
&
- G^{I J}_{U(t_i)}(\xii1+i0^+,\xii1-i0^+; \xii2+i0^+,\xii2-i0^+; \alpha_{t_i})
.
\label{DeltaNM}
\end{align}

Notice that
this analytic continuation is performed only on
the operators and not on the possible polarizations and momenta which are
still the ones allowed in the original region.

\subsection{Amplitudes with chiral vertices and plain twist fields}
In this section we would like to compute the correlators of $L_c$
chiral vertices as a warm up for the next section where we compute the
correlators of $N$ excited twists.
In particular we want to compute the generating function of the amplitudes
\begin{equation}
\langle  
\prod_{c=1}^{L_c}
\left[
\prod_{n=1}^\infty 
\left(\partial^n_u X^z \right)^{N_{(c) n}}
\left(\partial^n_u X^\bz \right)^{\bar N_{(c) n}}
\right](u_c) 
\prod_{t=1}^N \sigma_{\epsilon_t, f_t}(x_t)
\rangle
.
\label{sample-L-chiral-correlator}
\end{equation}
As in the previous section the SDS vertex for the emission of chiral
untwisted states from a twisted string with ends on $D_t$ and $D_N$ is given by
\begin{align}
\cS_{T(D_{t} D_N)}(u; \{c_{(c)}\})
=
&
\lim_{\eta \rightarrow 0^+}
e^{
\sum_{n=1}^\infty  
c_{(c) n I}~\partial^n_u 
\left[ 
X^{I(+)}_{T(D_{t} D_N)}(u, \bu)
+
X^{I(-)}_{T(D_{t} D_N)}(u e^{-\eta}, \bu e^{-\eta})
\right]
}
\nonumber\\
\times
&
\exp\left\{
-
\sum_{n,m=1}^\infty  
\bc_{(c) n }~c_{(c) m } ~\partial^n_{u_1} ~\partial^m_{u_2}
G_{U(t)}^{z \bz}(u_1, \bu_1; u_2, \bu_2; \alpha_t )
|_{u_1= u_2 e^{\eta} =u} 
\right\}
,
\label{S_chir_non_norm_ord0}
\end{align}
where we used the fact that 
$\partial^n_{u_1} ~\partial^m_{u_2} G_{U(t)}^{I J}$
is different from zero only when $I J= z \bz$ or $I J=\bz z$.
Moreover $G_{U(t)}^{z \bz}$ is actually independent on $t$ since it
does not depend on the phase $\alpha_t$.
The previous equation can be written in a more compact way 
by normal ordering the operatorial part and performing the limit. 
It then reads as
\begin{align}
\cS_{T(D_{t} D_N)}(u; \{c_{(c)}\})
=
&
:
e^{
\sum_{n=1}^\infty  
c_{(c) n I}~\partial^n_u 
X^{I}_{T(D_{t} D_N)}(u, \bu)
}
:
\nonumber\\
\times
&
\exp\left\{
\sum_{n,m=1}^\infty  
\bc_{(c) n }~c_{(c) m } ~\partial^n_{u_1} ~\partial^m_{u_2}
\Delta_{T(D_t D_N), chir}^{z \bz}(u_1, \bu_1; u_2, \bu_2; \PaMoSpNDuenn{t}{N} )
|_{u_1= u_2  =u} 
\right\}
,
\label{S_chir_non_norm_ord}
\end{align}
where we have introduced the derivative of the regularized 
chiral Green function as
\begin{align}
\partial_u \partial_v 
\Delta_{T(D_t D_N), chir}^{z \bz}(u, \bu; v, \bv; \PaMoSpNDuenn{t}{N})
&=
\partial_u \partial_v 
\Big[ G_{T(D_t D_N)}^{z \bz}(u, \bu; v, \bv; \PaMoSpNDuenn{t}{N})
\nonumber\\
&-
G^{z \bz}_{U(t)}(u, \bu; v, \bv; \alpha_t)
\Big]
.
\end{align}

Performing the same steps as in the previous section we get
\begin{align}
V_{N+L_c}(\{u_c, \{c_{(c)}\} \})
&=
\langle 
\sigma_{\epsilon_1,f_1}(x_1) \dots \sigma_{\epsilon_N,f_N}(x_N)
\rangle
\nonumber\\
\times
&
\prod_{c=1}^{L_c}
\Bigg\{
e^{
\sum_{n=1}^\infty c_{(c) n I }
\partial^{n-1}_{u_c} [ \partial_u X^I_{cl}(u_c, \bu_c; \PaMoSpFull) ]
}
\nonumber\\
\times
&
e^{
\oh
\sum_{n,m=1}^\infty c_{(c) n I } c_{(c) m J} 
\partial^{n}_{u_c} \partial^{m}_{v_{c}}
\Delta_{(N,M) (c)}^{I J}(u_c,\bu_c; v_c, \bv_c; \PaMoSp) |_{v_c=u_c}
]
}
\Bigg\}
\nonumber\\
\times
&
\prod_{1\le c < \hat c \le N}
e^{
\sum_{n,m=1}^\infty c_{(c) n I } c_{(\hat c) m J} 
\partial^{n}_{u_t} \partial^{m}_{ v_\htt}
G_{(N,M)}^{I J}(u_c, \bu_c; v_{\hat c}, \bv_{\hat c}  ; \PaMoSp)
}
,
\end{align}
where we have written the dependence on the complex conjugate
variables $\bu_c$ even if the derivatives of
$X^I_{cl}(u, \bu)$ and $G_{(N,M)}^{I J}(u, \bu; v, \bv  ; \PaMoSp)$ 
are independent of it.
We have done this in
order to be consistent with the notation used in the boundary case. 
The regularized chiral Green function is defined as expected as
\begin{align}
\partial_u \partial_v 
\Delta_{(N,M) (c)}^{I J}(u, \bu; v, \bv; \PaMoSp)
&=
\partial_u \partial_v 
\Big[ G_{(N,M)}^{I J}(u, \bu; v, \bv; \PaMoSp)
\nonumber\\
&-
G^{I J}_{U(t)}(u, \bu; v, \bv; \alpha_t)
\Big]
.
\label{regularized-chiral-Green}
\end{align}
Notice once more that the subtraction term is different from zero only
when $I J= z \bz$ or $I J= \bz z$ because of the derivatives.

There is actually an extra bonus in this approach:
we can also prove the explicit form of amplitudes involving closed
string states  that were conjectured in  \cite{Pesando:2014owa}.
This happens using the vertices for
the emission of closed string states written in open string formalism.
These vertices are written as a product of a chiral times an antichiral part (up
to cocycles).
Then we can apply the same procedure decribed before to these products. 


\subsection{Amplitudes with excited twist fields}
In this section we would like to compute the correlators of $N$
excited twists.
In particular we want to compute the generating function of the amplitudes
\begin{equation}
\langle  
\prod_{t=1}^N
\left[
\prod_{n=1}^\infty 
\left(\partial^n_u X^z \right)^{N_{n (t)}}
\left(\partial^n_u X^\bz \right)^{\bar N_{n (t)}}
\sigma_{\epsilon_t, f_t}
\right](x_t) \rangle
.
\label{sample-N-twist-correlator}
\end{equation}
The excited twists in the previous correlator can be described using
the operator to state correspondence.
According to the notation introduced in \cite{Pesando:2014owa} the
operator to state correspondence is given by
\begin{align}
\lim_{x\rightarrow 0}
&
\left[\prod_{n=0}^\infty 
\left(\partial^{n+1}_u X^z  \right)^{ N_n}
\left( \partial^{n+1}_u  X^\bz  \right)^{\bar N_n}
\sigma_{\epsilon_t, f_t}
\right](x)
|0\rangle_{SL(2)}=
\nonumber\\
&=
\prod_{n=0}^\infty 
\left(\kei{t} n! \alpha^\dagger_{n+\epsilon_t}\right)^{N_n}
\left(\kbei{t} n! \bar\alpha^\dagger_{n+\bar\epsilon_t} \right)^{\bar N_n}
|T_{T(D_{t-1}\, D_{t})}\rangle
\end{align}
with $\kei{t}= -i\oh\sqrt{2\alpha'} e^{i \pi \alpha_t}$ and
$\kbei{t}= -i\oh\sqrt{2\alpha'} e^{-i \pi \alpha_t}$\footnote{
There is obvioulsy also an analogous expression with the antichiral operators.
\COMMENTOO{ 
The limit is toward $0^+$, were it $0^-$ the two constants
$\ke$,$\kbe$ would be 
$\ke= -i\oh\sqrt{2\alpha'} e^{i \pi \alpha_{t-1}}$ and
$\kbe= -i\oh\sqrt{2\alpha'} e^{-i \pi \alpha_{t-1}}$.
}
}.
These states are built on the twisted vacuum 
$|T_{T(D_{t-1}\, D_{t})}\rangle
=\sigma_{\epsilon_t, f_t}(0)|0\rangle_{SL(2)} $ 
therefore the excited twists are naturally described in the
twisted Hilbert space $\cH_{T(D_{t-1}\, D_{t})}$.
However the $\cH_{T(D_{t-1} D_{t})}$ space is generically
not any of the twisted Hilbert spaces there are during the string
propagation.
They are $\cH_{T(D_t\, D_{N})}$ for $t=1, \dots N-1$ since
the $x<0$ ($\sigma=\pi$ in Minkowskian version) boundary of the string
is always attached on $D_N$.
This is luckily not a problem since we can describe any excited twist
as a limit of the product of a chiral operator and a plain twist
field, for example
\begin{align}
\Bigg(
&
\partial_u^{n-1} \left[ (u-x_t)^{\bep_t} \partial_u  X^z \right]~
\partial_u^{m-1} \left[ (u-x_t)^{\epsilon_t} \partial_u X^\bz \right]
\nonumber\\
&-
\partial_u^{n-1}  \partial_v^{m-1} 
  \left[ (u-x_t)^{\bep_t}  (v-x_t)^{\epsilon_t}~ 
        \partial_u \partial_v \Delta^{z \bz}_{T(D_{t-1} D_{t}),
          chir}(u-x_t; v-x_t;  \PaMoSpNDuenn{t-1}{t})
  \right]
\Bigg)
\sigma_{\epsilon_t, f_t} (x_t)
\nonumber\\
&=
\left( 
\partial^n X^z \partial^m X^\bz
\sigma_{\epsilon_t, f_t}
\right)(x_t)
+O(u-x_t)
.
\label{example-excited-twist-from-OPE}
\end{align}
In the previous expression the term with $ \partial_u \partial_v
\Delta^{z \bz}_{T(D_{t-1} D_{t}), chir}$  is necessary to cancel the
term which arises from the equation analogous to
eq. (\ref{S_chir_non_norm_ord}) for the $\cH_{T(D_{t-1}\, D_{t})}$
twisted Hilbert space.

In our case the plain twist field $\sigma_{\epsilon_t, f_t}(x_t)$ is hidden
into the boundary conditions.
We can get any excited twist field at $x_t$ by choosing the
appropriate chiral operator at $u$ and then take the limit
$u\rightarrow x_t$.
It follows then that it is enough to represent the abstract chiral operator
needed to create the wanted excited twisted field in the
desired twisted Hilbert space which can be either $\cH_{T(D_{t}\, D_{N})}$
or $\cH_{T(D_{t-1}\, D_{N})}$.

As discussed in  \cite{Pesando:2014owa} 
the generating function for the abstract
operators which is needed for generating any excited twists is given by
\begin{align}
\cT_{abs (t)}(u; \{ d_{(t)} \})
=
&
\exp\left\{
\sum_{n=1}^\infty  
\left[ 
\bar d_{(t) n } 
\partial_u^{n-1} \left[ (u-x_t)^{\bep_t} \partial_u X^z(u,\bu)\right]
+ d_{ (t) n} 
\partial_u^{n-1} \left[ (u-x_t)^{\epsilon_t} \partial_u X^\bz(u,\bu)\right]
\right]
\right\}
\nonumber\\
\exp
&\Bigg\{
-
\sum_{n,m=1}^\infty  
\bar d_{(t) n}~d_{(t) m} ~\partial^{n-1}_u ~\partial^{m-1}_v
\Big[ (u-x_t)^{\bep_t}  (v-x_t)^{\epsilon_t}
\nonumber\\
\times
&
\phantom{\exp\Big\{
\sum_{n=1}^\infty} 
\partial_u \partial_v \Delta^{z\bz}_{T(D_{t-1}\, D_t), chir}(u-x_t,
\bu-x_t; v-x_t, \bv-x_t; \PaMoSpNDuenn{t}{N})
\Big]
|_{v=u}
\Bigg\}
.
\label{T-abs}
\end{align}
Explicitly this means that
\begin{align}
\left[
\prod_{n=1}^\infty 
\left(\partial^n_u X^z\right)^{N_n}
\left(\partial^n_u X^\bz \right)^{\bar N_n}
\sigma_{\epsilon_t, f_t}
\right](x_t)
=
\lim_{u\rightarrow x_t}
\left.
\prod_{n=1}^\infty 
\frac{\partial^{N_n}}{\partial \bar d_{(t) n}^{N_n}}
\frac{\partial^{\bar N_n}}{\partial d_{(t) n}^{\bar N_n}}
\cT_{abs}(u; \{ d_{(t) }\}) \right|_{d=0}
\sigma_{\epsilon_t, f_t}(x_t)
.
\end{align}

The abstract generator (\ref{T-abs}) can be realized in any Hilbert
space.
In particular in the twisted Hilbert space 
$\cH_{T(D_t\,  D_{N})}$  where $x_t<|u|<x_{t-1}$ it reads
\begin{align}
\cT_{T(D_{t}\, D_N)}(u; \{d_{(t)}\})
=
&:
\exp\left\{
\sum_{n=1}^\infty  
d_{(t) n I}~\partial^{n-1}_u \left[ (u-x_t)^{\epsilon_{ t~ I}}
  \partial_u X^I_{T(D_{t}\, D_N)}(u) \right]
\right\} 
:
%
\nonumber\\
\exp
&\Bigg\{
\sum_{n,m=1}^\infty  
\bar d_{(t) n}~d_{(t) m} ~\partial^{n-1}_u ~\partial^{m-1}_v
\Big[ (u-x_t)^{\bar \epsilon_t}  (v-x_t)^{\epsilon_t} 
\nonumber\\
&
\phantom{\exp\Big\{
\sum_{n=1}^\infty} 
\partial_u \partial_v \Delta^{z\bz}_{T(D_t\, D_N), chir}(u,\bu; v,\bv; \PaMoSpNDuenn{t}{N})
\Big]
\Big|_{v=u}
\Bigg\}
.
\nonumber\\
%
\times
\exp
&\Bigg\{
-
\sum_{n,m=1}^\infty  
\bar d_{(t) n}~d_{(t) m} ~\partial^{n-1}_u ~\partial^{m-1}_v
\Big[ (u-x_t)^{\bep_t}  (v-x_t)^{\epsilon_t}
\nonumber\\
&
\phantom{\exp\Big\{
\sum_{n=1}^\infty} 
\partial_u \partial_v \Delta^{z\bz}_{T(D_{t-1}\, D_t), chir}(u-x_t,
\bu-x_t; v-x_t, \bv-x_t; \PaMoSpNDuenn{t-1}{t})
\Big]
\Big|_{v=u}
\Bigg\}
.
\label{T_bou}
\end{align}
The first two factors on the right hand side are nothing else but the
chiral SDS vertex with $u$ dependent $c_{n I}$, roughly
$c_{n I} \rightarrow d_{n I} (u-x_t)^{\epsilon_{ t~ I}}$.
Therefore the $\Delta^{z \bz}$ in
the quadratic term depends on $u$ and $v$ only and not on $x_t$.
The last factor is the ``normalization'' of the abstract vertex,
i.e. the second factor on the right hand side in eq. (\ref{T-abs}) and
therefore depends on $u-x_t$ and $v-x_t$.

It is now enough to insert a SDS for any  untwisted operator into the
radial ordered  expression
\begin{align}
V_{0+N}( \{x_t;\{d_{(t)}\}\})=
\lim_{ \{u_t\} \rightarrow \{x_t\} }
\langle 0_{out}| 
R\left[ \prod_{t=1}^N \cT_{T(D_{t} D_N)}(u_t; \{d_{(t)}\})  \right]
| 0_{in}\rangle
\label{N-twisted-Reggeon-restricted}
\end{align}
in order to compute the generating function of all the correlators
like (\ref{sample-N-twist-correlator}).

The computation of the expectation value is the same as before for the
chiral vertices.
What is interesting is to trace the contributions to self
interactions.
For any $t$ we get two contributions: one from the
expectation value and the other from the ``normalization'' of the
abstract vertex, i.e. the third factor in eq. (\ref{T_bou}).
We get therefore
\begin{align}
%
%
\exp
&\Bigg\{
\oh
\sum_{n,m=1}^\infty  
d_{(t) n I}~d_{(t) m J} ~\partial^{n-1}_u ~\partial^{m-1}_v
\Big[ (u-x_t)^{\epsilon_{t\, I}}  (v-x_t)^{\epsilon_{t\, J}} 
\partial_u \partial_v \Delta^{I J}_{(N, M) (c_t)}(u,\bu; v,\bv; \PaMoSpht)
\Big]
\Big|_{v=u=u_t}
\Bigg\}
\nonumber\\
%
\times
\exp
&\Bigg\{
-
\sum_{n,m=1}^\infty  
\bar d_{(t) n}~d_{(t) m} ~\partial^{n-1}_u ~\partial^{m-1}_v
\Big[ (u-x_t)^{\bar \epsilon}  (v-x_t)^{\epsilon}
\nonumber\\
& \phantom{\exp\Big\{ \sum_{n=1}^\infty \bar d_{(t) n}~d_{(t) m}} 
\partial_u \partial_v \Delta^{z\bz}_{chir, T(D_{t-1} D_t)}(u-x_t,
\bu-x_t; v-x_t, \bv-x_t; \PaMoSpNDuenn{t-1}{t})
\Big]
\Big|_{v=u=u_t}
\Bigg\}
,
\end{align}
where $c_t$ is the index associated with the chiral vertex at
$u_{c_t}=u_t$ and
$\partial_u \partial_v \Delta^{I J}_{(N, M) (c_t)}$ is the derivative
of  regularized Green function given in eq. (\ref{regularized-chiral-Green}).
This expression can be written in a more compact way as
\begin{align}
%
%
\exp
&\Bigg\{
\oh
\sum_{n,m=1}^\infty  
d_{(t) n I}~d_{(t) m J} ~\partial^{n-1}_u ~\partial^{m-1}_v
\Big[ (u-x_t)^{\epsilon_{t\, I}}  (v-x_t)^{\epsilon_{t\,J}} 
\partial_u \partial_v \Delta^{I J}_{(N, M) (t)}(u,\bu; v,\bv; \PaMoSpht)
\Big]
\Big|_{v=u=u_t}
\Bigg\}
,
\end{align}
where we have defined the (derivative of the) regularized Green
function at the position $x_t$ of the twist fields $t$ to be
\begin{align}
\partial_u \partial_v &\Delta_{(N,M) (t)}^{I J}(u, \bu; v, \bv; 
\PaMoSpht ) 
=
\nonumber\\
=
&
\partial_u \partial_v 
\Big[ 
\Delta_{(N,M) (c_t)}^{I J}(u, \bu; v, \bv; \PaMoSpht)
-
\Delta^{I J}_{chir, T(D_{t-1} D_t)}(u-x_t,
\bu-x_t; v-x_t, \bv-x_t; \PaMoSpNDuenn{t-1}{t})
\Big]
\nonumber\\
=
&
\partial_u \partial_v \left[ G_{(N,M)}^{I J}(u, \bu; v, \bv ; \PaMoSpht)
-
G_{N=2, T(D_t D_N)}^{I J}(u-x_t, \bu-x_t; v-x_t, \bv-x_t;
\PaMoSpNDuenn{t}{N})
\right]
\end{align}
and we used $G_{U(t)}^{I J}(u-x_t, v-x_t) =G_{U(t)}^{I J}(u, v)$
to write the last line.
Notice that when $x_{N-1}=0$ the previous equation 
becomes eq. (\ref{DeltaNM-t=N-1}).
Actually because of the chiral derivatives the previous expression
simplifies fot two combintations of indeces $I J$ to
\begin{align}
\partial_u \partial_v \Delta_{(N,M) (t)}^{z z}(u, \bu; v, \bv ; \PaMoSpht)
&= 
\partial_u \partial_v G_{(N,M)}^{z z}(u, \bu; v, \bv; \PaMoSpht)
\nonumber\\
\partial_u \partial_v \Delta_{(N,M) (t)}^{\bz \bz}(u, \bu; v, \bv; \PaMoSpht)
&=
\partial_u \partial_v G_{(N,M)}^{\bz \bz}(u, \bu; v, \bv; \PaMoSpht)
\end{align}
The third combination  corresponds
to $\partial_u \partial_v \Delta_{(N,M) (t)}^{z \bz}$ and does not
simplify.

On general basis the regularized Green function is obtained by subtracting 
the divergent part with the proper monodromy at the
point of regularization.   
At the point where a twist field is located the divergent part with
the proper monodromy  means $G_{N=2, T(D_{t-1}\, D_t)}$ 
while in all other points means $G_U$.
In particular both 
$(u-x_t)^{\epsilon_{t I}} (v-x_t)^{\epsilon_{t J}}
\partial_u \partial_v  \Delta_{(N,M) (t)}^{I J}$
and
$
(u-x_t)^{\epsilon_{t I}} (v-x_t)^{\epsilon_{t J}} \partial_u \partial_v  G_{(N,M)}^{I J}
$
are analytic functions at $u=x_t$. 

Assembling all pieces we get therefore the generating function for the
excited twists correlator
\begin{align}
V_{0+N}&( \{x_t;\{d_{(t)}\}\})
=
\lim_{\{u_t\} \rightarrow \{x_t\} }
\langle 
\sigma_{\epsilon_1,f_1}(x_1) \dots \sigma_{\epsilon_N,f_N}(x_N)
\rangle
\nonumber\\
\times
&
\prod_{t=1}^N
\Bigg\{
e^{
\sum_{n=1}^\infty d_{(t) n I }
\partial^{n-1}_{u_t} [ (u_t-x_t)^{\epsilon_{t I}} \partial_u
  X^I_{cl}(u_t, \bu_t; \PaMoSpFull) ]
}
\nonumber\\
\times
&
e^{
\oh
\sum_{n,m=1}^\infty d_{(t) n I } d_{(t) m J} 
\partial^{n-1}_{u_t} \partial^{m-1}_{v_t}
[(u_t-x_t)^{\epsilon_{t I}} (v_t-x_t)^{\epsilon_{t J}}
\partial_u \partial_v 
\Delta_{(N,M) (t)}^{I J}(u_t, \bu_t; v_t, \bv_t; \PaMoSpht) ]
|_{v_t=u_t}
}
\Bigg\}
\nonumber\\
\times
&
\prod_{1\le t < \htt \le N}
e^{
\sum_{n,m=1}^\infty d_{(t) n I } d_{(\htt) m J} 
\partial^{n-1}_{u_t} \partial^{m-1}_{ v_\htt}
[(u_t-x_t)^{\epsilon_{t I}} (v_\htt-x_\htt)^{\epsilon_{\htt J}}
\partial_u \partial_v
G_{(N,M)}^{I J}(u_t, \bu_t; v_\htt, \bv_\htt  ; \PaMoSpht)
]
}
.
\label{reggeon-excited-twists}
\end{align}

\subsection{The generating function for for $N$ excited twist fields and $L$ plain vertices}
We are now in the position of computing the desired generating
function for $N$ excited twist fields and $L$ plain vertices.
It simply amounts to the computation of
\begin{align}
V_{L+N}( \{x_t;\{d_{(t)}\}\})=
\lim_{ \{u_t\} \rightarrow \{x_t\} }
\langle 0_{out}| 
R\left[ 
\prod_{i=1}^L \cS_{T(D_{t} D_N)}(\hx_i; \{c_{(i)}\})  
\prod_{t=1}^N \cT_{T(D_{t} D_N)}(u_t; \{d_{(t)}\})  
\right]
| 0_{in}\rangle
.
\label{N-twisted-Reggeon-restricted}
\end{align}
This computation can be done as explained in the previous sections.
The result is made of the product of three blocks: 
interactions between two excited twists,
interactions between two plain vertices and interactions between
one excited twists and one plain vertex.
This structure is evident in the final result 
\begin{align}
V_{N+L}(K_t,J_i)
&=
\lim_{\{u_t\} \rightarrow \{x_t\}}
\langle 
\sigma_{\epsilon_1,f_1}(x_1) \dots \sigma_{\epsilon_N,f_N}(x_N)
\rangle
\nonumber\\
\times
&
\prod_{t=1}^N
\Bigg\{
e^{
\sum_{n=1}^\infty d_{(t) n I }
\partial^{n-1}_{u_t} [ (u_t-x_t)^{\epsilon_{t I}} \partial_u
  X^I_{cl}(u_t, \bu_t; \PaMoSpFull) ]
}
\nonumber\\
\times
&
e^{
\oh
\sum_{n,m=1}^\infty d_{(t) n I } d_{(t) m J} 
\partial^{n-1}_{u_t} \partial^{m-1}_{v_t}
[(u_t-x_t)^{\epsilon_{t I}} (v_t-x_t)^{\epsilon_{t J}}
\partial_u \partial_v 
\Delta_{(N,M) (t)}^{I J}(u_t, \bu_t; v_t, \bv_t; \PaMoSpht) ]
|_{v_t=u_t}
}
\Bigg\}
\nonumber\\
\times
&
\prod_{i=1}^L
\Bigg\{
e^{
\sum_{n=0}^\infty c_{(i) n I}
\partial^n_{x_i} X^I_{cl}(x_i, x_i; \PaMoSpFull)
}
\nonumber\\
\times
&
e^{
\oh
\sum_{n=0}^\infty c_{(i) n I}
\sum_{m=0}^\infty c_{(i) m J} 
\partial^n_{x_i} \partial^m_{ \hat x_i}
\Delta^{I J}_{(N,M), bou (i)}(x_i, \hat x_i ; \PaMoSpht) |_{ \hat x_i= x_i}
}
\Bigg\}
\nonumber\\
\times
&
\prod_{1\le t < \htt \le N}
e^{
\sum_{n,m=1}^\infty d_{(t) n I } d_{(\htt) m J} 
\partial^{n-1}_{u_t} \partial^{m-1}_{ v_\htt}
[(u_t-x_t)^{\epsilon_{t I}} (v_\htt-x_\htt)^{\epsilon_{\htt J}}
\partial_u \partial_v
G_{(N,M)}^{I J}(u_t, \bu_t; v_\htt, \bv_\htt  ; \PaMoSpht)
]
}
\nonumber\\
\times
&
\prod_{1\le i < j\le L}
e^{
\sum_{n=0}^\infty c_{(i) n I}
\sum_{m=0}^\infty c_{(j) m J} 
\partial^n_{x_i} \partial^m_{x_j}
G^{I J}_{(N,M), bou}(x_i, x_j ; \PaMoSpht)
}
\nonumber\\
\times
&
\prod_{1\le t \le N}
\prod_{1\le j \le L}
e^{
\sum_{n=1}^\infty d_{(t) n I } c_{(j) m J} 
\partial^{n-1}_{u_t} \partial^{m}_{x_j}
[(u_t-x_t)^{\epsilon_{t I}} \partial_u 
G_{(N,M)}^{I J}(u_t, \bu_t; x_j, x_j  ; \PaMoSpht)
]
}
,
\label{reggeon-excited-twists+bou}
\end{align}
where the last line is exactly due to the interactions between one
excited twist and one plain vertex.

\appendix

\section{Self-adjointness of the laplacian}
\label{app:self-adjoint}

We want to show that $\partial\bar\partial$ is a
self-adjoint operator only if we use the quantum boundary conditions.
In particular we define
$\partial\bar\partial=\partial_x^2+\partial_y^2$ 
as operator which acts
on a couple of complex functions $f^I(u,\bu)$ defined on the upper half plane.
Then we take not only $f^I\in L^2(H)$ but we require that 
$\partial_x f^I$, $\partial_y f^I$, $\partial_x^2 f^I$ and
$\partial_y^2 f^I$ be defined almost everywhere and  that the action
$ \int_H d x ~ d y~ f^{I *} (\partial_x^2+\partial_y^2)f^I$ be finite.
Since we need to integrate by part  we need
\begin{equation}
\int_a^b d x \partial_x^2 f^I(u,\bu) =
\partial_x f^I(b+i y,b-i y ) - \partial_x f^I(a+i y,a-i y ) 
\end{equation}
(and similarly for $y$) hence $\partial_x f^I$ and $\partial_y f^I$  
must be absolutely continuous.
The similar condition with a single derivative 
is a consequence of the existence almost everywhere of $\partial_x^2 f^I$,
$\partial_y^2 f^I$ which imply that $\partial_x f^I$, $\partial_y f^I$
be almost everywhere continuous.

Finally we impose the boundary conditions
\begin{equation}
f^z(x,x)=e^{i 2\pi \alpha_t} f^\bz(x,x),~~~~
\partial_y f^z(x,x)= -e^{i 2\pi \alpha_t} \partial_y f^\bz(x,x),~~~~
x\in(x_{t},x_{t-1})
\end{equation}
and
\begin{equation}
f^I(u,\bu)\rightarrow 0 ~\mbox{as}~ u\rightarrow \infty
.
\end{equation}
Now we can determine the domain of the dual operator, i.e. we
determine the conditions we must impose on an arbitrary vector $g^I$ so
that we can write 
$(g, \partial\bar\partial f)=(\partial\bar\partial g, f)$.
In order to do so we compute using the previous boundary conditions
\begin{align}
&\int_H d x~d y~ g^{I*} (\partial_x^2+\partial_y^2)f^I
\nonumber\\
=&
\int_0^\infty d y 
[  g^{z*} \partial_x f^{z} + g^{\bz *} \partial_x f^{\bz} ]
|^{x=+\infty}_{x=-\infty}
\nonumber\\
&+
\sum \int_{x_{t}}^{x_{t-1}} d x~ 
 [g^z(x,x)-e^{i 2\pi \alpha_t} g^\bz(x,x)]^* \partial_y f^z
\nonumber\\
&+
\sum \int_{x_{t}}^{x_{t-1}} d x~ 
 [\partial_y g^z(x,x)+e^{i 2\pi \alpha_t} \partial_y g^\bz(x,x)]^*  f^z
\nonumber\\
&+
\int_H d x~d y~(\partial_x^2+\partial_y^2)g^{I *} f^I
\end{align}
from which we see that $g^I$ must satisfy the same boundary conditions
as $f^I$ and hence the operator is not only Hermitian but self-adjoint.

\COMMENTOO{
The boundary conditions imply that
\begin{equation}
\partial f^z \sim (u-x)^{-\epsilon+n},~~~~
\partial f^\bz \sim (u-x)^{-\bar\epsilon+\bar n},~~~~
\end{equation}
and the finiteness of the action requires then $n,\bar n \ge 0$ and therefore
\begin{equation}
f^z(x,x)=f^\bz(x,x)=0
\end{equation}.
Actually the request of $\partial\partial f$ belonging to $L^2$
requires $n,\bar n \ge 1$.
} 

\section{Details on the metric for modes}
\label{app:details_metric}
Consider for example the computation $(\bX_{(a) 1}, \bX_{(a)2 })$.
In the following we write $\cbX_1= \bar\cG$ and $\cbX_2= \bar\cF$ 
for notational simplicity.
It is immediate to get
\begin{align}
-i~\int_{|u|=r_0} *j
&=
r_0
\int_0^\pi d \theta~ \Big[
 \left(\bar \cG (r e^{i \theta})\right)^*   \partial_r\bar \cF (r e^{i \theta})
+ \left(\bar \cG (r e^{-i \theta})\right)^*   \partial_r\bar \cF (r e^{-i \theta})
\nonumber\\
&
-\partial_r\left( \bar \cG (r e^{i \theta}) \right)^* \bar \cF (r e^{i \theta})
-\partial_r\left( \bar \cG (r e^{-i \theta}) \right)^* \bar \cF (r e^{-i \theta})
\Big]\Big|_{r=r_0}
-
\end{align} 
Now we rewrite 
$\partial_r\bar \cF (r e^{i \theta})= -\frac{i}{r} \partial_\theta\bar
\cF (r e^{i \theta}) $ and so on for all the other derivatives then we
get
\begin{align}
-i~\int_{|u|=r_0} *j
&=
i
\int_0^\pi d \theta~ \Big[
\partial_\theta \left(
 \left(\bar \cG (r_0 e^{-i \theta})\right)^*  \bar \cF (r_0 e^{-i \theta})
\right)
-
\partial_\theta \left(
\left( \bar \cG (r_0 e^{+i \theta})\right)^*  \bar \cF (r_0 e^{+i \theta})
\right)
\Big]
\end{align} 
which vanishes because of the boundary conditions which ensure that
$
\left( \bar \cG (r_0 e^{i \theta})\right)^*  
\bar \cF (r_0 e^{i \theta}) \Big|_{\theta=0^+}
=
\left( \bar \cG (r_0 e^{-i \theta})\right)^*  
\bar \cF (r_0 e^{-i \theta}) \Big|_{\theta=0^+}
$
.
Similarly for $(\bX_{(a)}, X_{(c)})$ we arrive  to
\begin{align}
-i~\int_{|u|=r_0} *j
&=
i e^{-i 2\pi \alpha_1}
\int_{-\pi}^\pi d \theta~ \Big[
\left(\partial_\theta \bar \cG (r_0 e^{-i \theta})\right)^*  \cF (r_0 e^{i \theta})
-
\left(\bar \cG (r_0 e^{-i \theta})\right)^*  \partial_\theta\cF (r_0 e^{+i \theta})
\Big]
,
\end{align} 
where it is meaningful write the integration interval as $[-\pi, \pi]$
since both the terms are continuous at $\theta=0$, i.e. for example
$
\left(\partial_\theta \bar \cG (r_0 e^{-i \theta})\right)^*  
\cF(r_0 e^{+i \theta}) |_{\theta=0^+}
=
\left(\partial_\theta \bar \cG (r_0 e^{i \theta})\right)^*  
\cF (r_0 e^{-i \theta})
|_{\theta=0^+}
$.
Now integrating by part the first and using the boundary conditions to
evaluate to zero the constant obtained from the integration by part
we find
\begin{align}
-i~\int_{|u|=r_0} *j
&=
-2 i e^{-i 2\pi \alpha_1}
\int_{-\pi}^\pi d \theta~ 
\left(\partial_\theta \bar \cG (r_0 e^{-i \theta})\right)^*  
\cF (r_0 e^{+i \theta})
\end{align}



\end{document}